\documentclass[fleqn,usenatbib]{mnras}

\usepackage{newtxtext,newtxmath}
\usepackage{xcolor}
\usepackage[normalem]{ulem}
\usepackage[T1]{fontenc}

\DeclareRobustCommand{\VAN}[3]{#2}
\let\VANthebibliography\thebibliography
\def\thebibliography{\DeclareRobustCommand{\VAN}[3]{##3}\VANthebibliography}

\usepackage{graphicx}	% Including figure files
\usepackage{amsmath}	% Advanced maths commands
\usepackage{multirow}
\newcommand{\cm}{\,{\rm cm}\,}% centimeters
\newcommand{\pc}{\,{\rm pc}\,}% parsec
\newcommand{\kpc}{\,{\rm kpc}\,}% kiloparsec
\newcommand{\Myr}{\,{\rm Myr}\,}% megayears
\newcommand{\Gyr}{\,{\rm Gyr}\,}% gigayears
\newcommand{\kms}{\,{\rm km s}$^{-1}$\,}% kilometres per second
\newcommand{\G}{\,{\rm G}\,}% gauss
\newcommand{\muG}{\,$\mu$\G\,}% microgauss
\newcommand{\g}{\,{\rm g}\,}% gram
\newcommand{\eref}[1]{Eq.~(\ref{#1})}
\newcommand{\fref}{Fig.\,\ref}% referring figures
\newcommand{\sref}{Sec.\,\ref}% referring sections
\title[Observational Signatures of Galactic Turbulent Dynamos]{Observational Signatures of Galactic Turbulent Dynamos}

\author[Yann Carteret et al.]{
Yann Carteret$^{1}$\thanks{E-mail:yann.carteret@gmail.com},
Abhĳit B. Bendre$^{1}$, and 
Jennifer Schober$^{1}$
\\
$^{1}$Laboratoire d'Astrophysique, EPFL, CH-1290 Sauverny, Switzerland\\
}

\date{Accepted XXX. Received YYY; in original form ZZZ}

\pubyear{2022}

\begin{document}
\label{firstpage}
\pagerange{\pageref{firstpage}--\pageref{lastpage}}
\maketitle

\begin{abstract}
We analyse the observational signatures of galactic 
magnetic fields that are self-consistently generated 
in magnetohydrodynamic simulations of the interstellar 
medium through turbulence driven by supernova (SN) 
explosions and differential rotation. In particular, 
we study the time evolution of the Faraday rotation 
measure (RM), synchrotron radiation, and Stokes 
parameters by characterising the typical structures 
formed in the plane of observation. We do this by 
defining two distinct models for both thermal and 
cosmic ray (CR) electron distributions. Our results 
indicate that the maps of RM have structures which 
are sheared and rendered anisotropically by differential 
%JS: added "the"
%rotation and that they depend on the choice of thermal 
rotation and that they depend on the choice of the thermal 
electrons model as well as the SN rate. Synchrotron 
maps are qualitatively similar to the maps of the
mean magnetic field along the line of sight and 
structures are only marginally affected by the CR 
model. Stokes parameters and related quantities, such 
as the degree of linear polarisation, are highly 
dependent on both frequency and resolution of the 
observation. 
\end{abstract}

\begin{keywords}
ISM: magnetic fields -- radio continuum: ISM -- galaxies: ISM -- galaxies: magnetic fields -- (magnetohydrodynamics) MHD -- methods: data analysis
\end{keywords}

%%%%%%%%%%%%%%%%%%%%%%%%%%%%%%%%%%%%%%%%%%%%%%%%%%

%%%%%%%%%%%%%%%%% BODY OF PAPER %%%%%%%%%%%%%%%%%%
\section{Introduction}
Observations of the polarised radio synchrotron radiation 
The polarised radio synchrotron radiation 
emitted from the interstellar medium (ISM) of nearby disc 
galaxies and its associated Faraday rotation probe the 
topology and strength of the magnetic field hosted by them. 
The strength of the component along the line of sight (LOS) 
of the observer is estimated from the Faraday rotation 
measure (RM), while from the intensity of the radiation 
and its angle of polarisation the strength of the planar 
field component could be inferred 
\citep[e.g.][and references therein]{BeWi}. 
As such, it has been established that regular magnetic 
fields of kilo-parsec scales in coherence as well as the 
irregular or small-scale magnetic fields are abundantly 
present in nearby 
\citep[e.g.][]{sofue1986global,thompson2006magnetic,krause2008magnetic,fletcher} 
and also in high-redshift disc galaxies 
\citep[e.g.][]{Bernet2008}. 
Typical strengths of these fields 
are about a few tens of a $\mu$G in Milky Way like galaxies \citep{Beck_2004},
which corresponds to approximate equipartition between magnetic and
turbulent kinetic energy. 

The mechanism of the growth and 
sustenance of large-scale magnetic fields in galaxies, although unclear, is very likely a turbulent dynamo, operating 
in their ISM, as a result of an induction effect generated 
from a combined action of supernova (SN) driven turbulence,
differential rotation, and density stratification 
\citep{radler1969,Parker,radler1980,ZeldovichBook,Brandenburg2005}. 
This process explains the amplification of very weak 
initial fields 
%JS: u -> U
%(which might have been generated in the early universe
(which might have been generated in the early Universe
\citep{KandusEtAl2011, Subramanian2016} or through 
astrophysical mechanisms
\citep{Biermann1950,subramanian1994thermal,KulsrudEtAl1997})
to a strength roughly in equipartition 
with the turbulent kinetic energy density. It has also been 
demonstrated in several empirical studies 
\citep[see e.g.][and references therein]{shukurov2006galactic} 
through turbulent 
transport mechanisms, as well as in direct numerical 
studies with varying setups 
\citep[see eg.][etc.]{gressel2008direct,hanasz2009global,gent2013supernova}. 

A better understanding of cosmic magnetic fields 
and mechanisms of their origin would enhance a general 
comprehension in numerous domains of astrophysics as they are believed 
to play an important role for example in the angular 
momentum transport in accretion discs \citep{shaku}, 
in the morphology of the ISM and the star 
formation process \citep{price,KrumholzFederrath2019}, 
and also in the propagation of cosmic rays \citep{yan}. 
Probing the observational signatures of the dynamo 
mechanism and being able to distinguish between 
various scenarios of generating the galactic magnetic 
field is therefore an important open problem. 
Scenarios of magnetic field evolution 
include different kinds of dynamos, 
like the small-scale or
fluctuation dynamo which is fast but only produces magnetic fields on length scales
smaller than the injection scales, the large-scale or mean-field dynamo, and
combinations of the two.

Complications in observing cosmic magnetism arise primarily  
because the observables depend not 
only upon the field itself but also on the 
spatial distribution of thermal and non-thermal electrons, as well as on 
their statistical correlations with the magnetic 
field, which are usually not well understood. Typically 
observational studies of galactic magnetic fields 
rely upon theoretically inferred models of these 
distributions, and it is thus important to understand 
the extent to which these observational inferences depend upon 
the models of electron distributions. 

In the current study, we address the aforementioned issue and 
estimate observables associated with galactic magnetic 
fields from numerical simulations of the mean-field 
dynamo with the goal of improving the interpretation 
of observations of real galaxies. Similar synthetic 
observational analyses have been performed for 
interpreting the magnetic fields in the young galaxies
\citep{bhat2013fluctuation,sur2018}, and also for the simulations of 
clusters of galaxies \citep{Sur2021}, where the 
small-scale or fluctuation dynamo is thought to be the
prevalent mechanism for field generation \citep{SchoberEtAl2013}. In the 
present analysis we focus on the ISM 
magnetic fields generated as a result of a large-scale 
dynamo. 

We base this study on the data of the 
MHD simulations of galactic ISM performed by 
\citet{Bendre2015}, the relevant details of which are 
also described in the following section. We process this 
data to perform a series of synthetic radio observations,
and use physically motivated models for the 
distributions of thermal and CR electrons which are also 
crucial for modelling the continuum radio emission 
spectrum. We further explore the dependence of its 
polarisation angle, the two point correlation function 
of radio emission etc. upon these distributions.
In \cite{RappazEtAL2022}, a similar analysis of ISM 
simulations has been presented. These were also based 
on the MHD simulations of a local patch of galactic 
ISM, stirred with SN driven turbulence, which focused 
on realistically simulating the multi-phase morphology 
of ISM along with a detailed chemical network, and not 
specifically on the dynamo mechanism itself. The 
distribution of magnetic fields in these simulations 
therefore was a result of asymptotic turbulent decay 
of initially imposed uniform magnetic field. The 
simulations that we use here however, were focused on 
self consistently generating the large-scale dynamo 
effect from direct simulations of SN driven turbulence 
and differential shear. 

The paper is organised as follows: In Sec.~\ref{s2} we summarise the 
basic setup and the key results of the galactic dynamo simulations. 
In Sec.~\ref{sec_observables}, we will present the post-processing of the 
simulations that allows us to extract various mock observables.
Telescope effects will be discussed in Sec.~\ref{tel_eff}. 
In Sec.~\ref{sec_discussion} we comment on our different assumptions
and we draw our conclusions in Sec.~\ref{sec_conclusions}.

\section{Simulations}
\label{s2}

In this section we briefly describe the setup of the 
direct numerical simulations (DNS) we 
analyse in 
the following sections. Detailed 
discussion of the DNS setup and its various outcomes are also presented in \citet{Bendre2015}.

These were non-ideal MHD simulations of a local 
box
of the ISM in a typical spiral galaxy, performed using 
\texttt{NIRVANA} MHD \citep{Nirvanacode} code. The simulation domain 
spanned $\sim$ 0.8 \kpc by 0.8 \kpc range in radial
($x$) and azimuthal ($y$) directions, while in the 
vertical $z$ direction it spanned $\sim$-2.1 to 2.1
\kpc above and below the galactic mid-plane. The 
domain was resolved in $96\times 96 \times 512$ 
cells amounting to an uniform Cartesian grid with a 
resolution of 
$\delta \sim 8$\pc. Shearing periodic 
boundary conditions were used at the radial while 
periodic ones were used at the $y$ boundaries, to 
simulate respectively the radial shear and axisymmetry
of azimuthal galactic flows. 
A flat rotation curve 
of galactic rotation was also included by letting the 
angular velocity decrease with radius $R$ as $\Omega
\propto1/R$, with $\Omega_0 = 100~$\kms\kpc$^{-1}$ at
the centre of the domain.
The simulated ISM was 
composed only of hydrogen, although its multi-phase 
morphology was still captured 
in a rudimentary way
by incorporating the temperature dependent rates of 
heat transfer, 
representing a piece-wise power law
of radiative cooling \citep[similar to][]{sachez_salcedo}. 
The initial mass density $\rho$ was vertically 
stratified and balanced hydrostatically with 
gravity. 
It had a scale height of $\sim 300$\pc 
(and midplane value of $\sim10^{-24}$\g \cm$^2$).
Outflow boundary conditions were used at the vertical 
boundaries which allowed the outflow of the matter but 
restricted its inflow.
SN explosions were simulated as spontaneous local 
expulsions of thermal energy injected at random 
locations scaling with the density and at predefined 
rates of 25\%, 50\%, and 100\% of the average Milky Way 
SN
rate 
\citep[$\sim30~$\Myr$^{-1}$ \kpc$^{-2}$, e.g.,][]{ferrier}. 
We analyse all of these three models 
and refer to these as Run25, Run50, and Run100 
respectively in the following sections. 

\begin{figure}
	\includegraphics[width=\columnwidth]{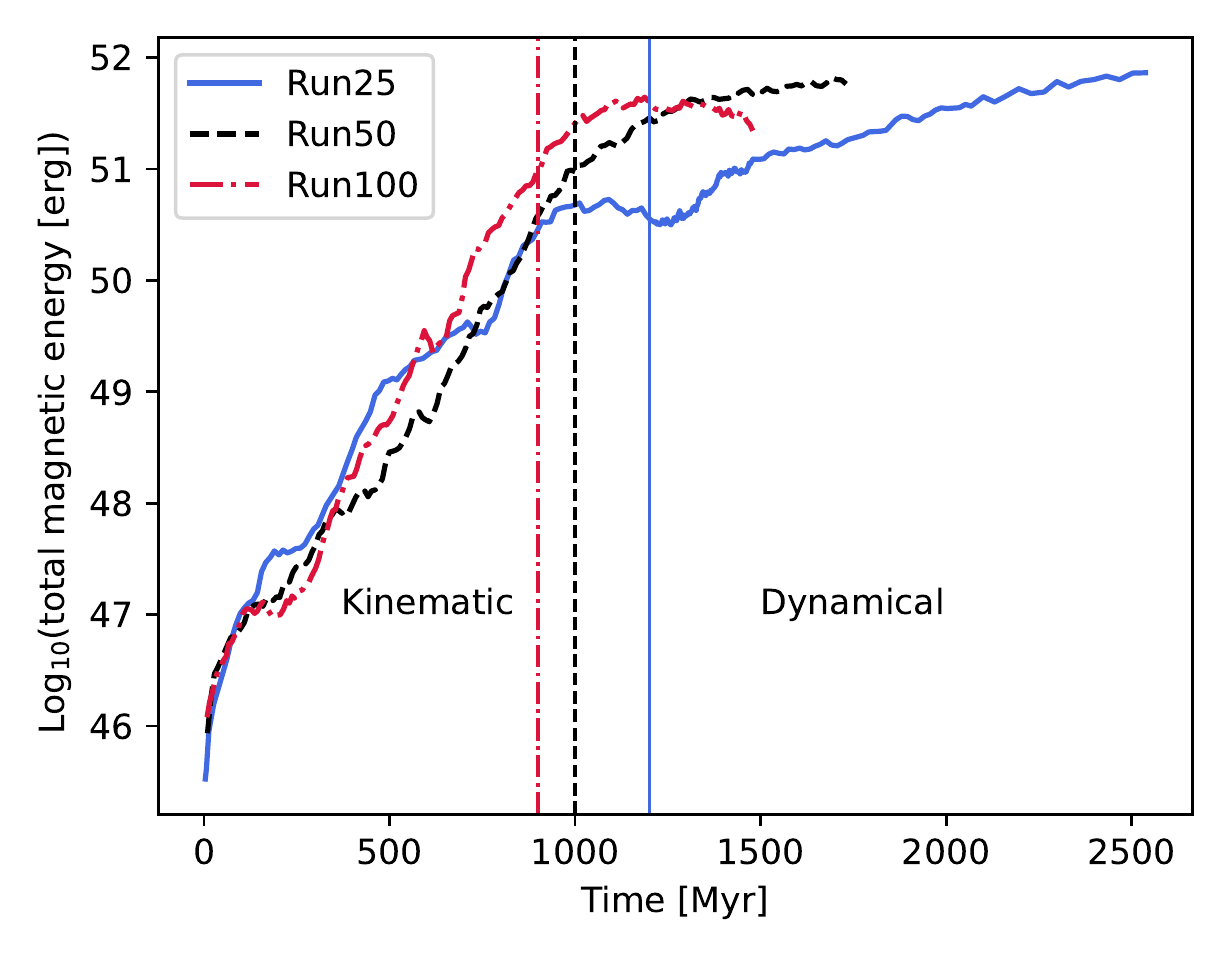}
    \caption{Time evolution of the magnetic field energy for
    the three runs. 
    The thin vertical lines indicate the transition from the 
    kinematic to the dynamical dynamo phase for the different runs.
    }
    \label{fig:Mag_energy}
\end{figure}

\begin{figure}
	\includegraphics[width=\columnwidth]{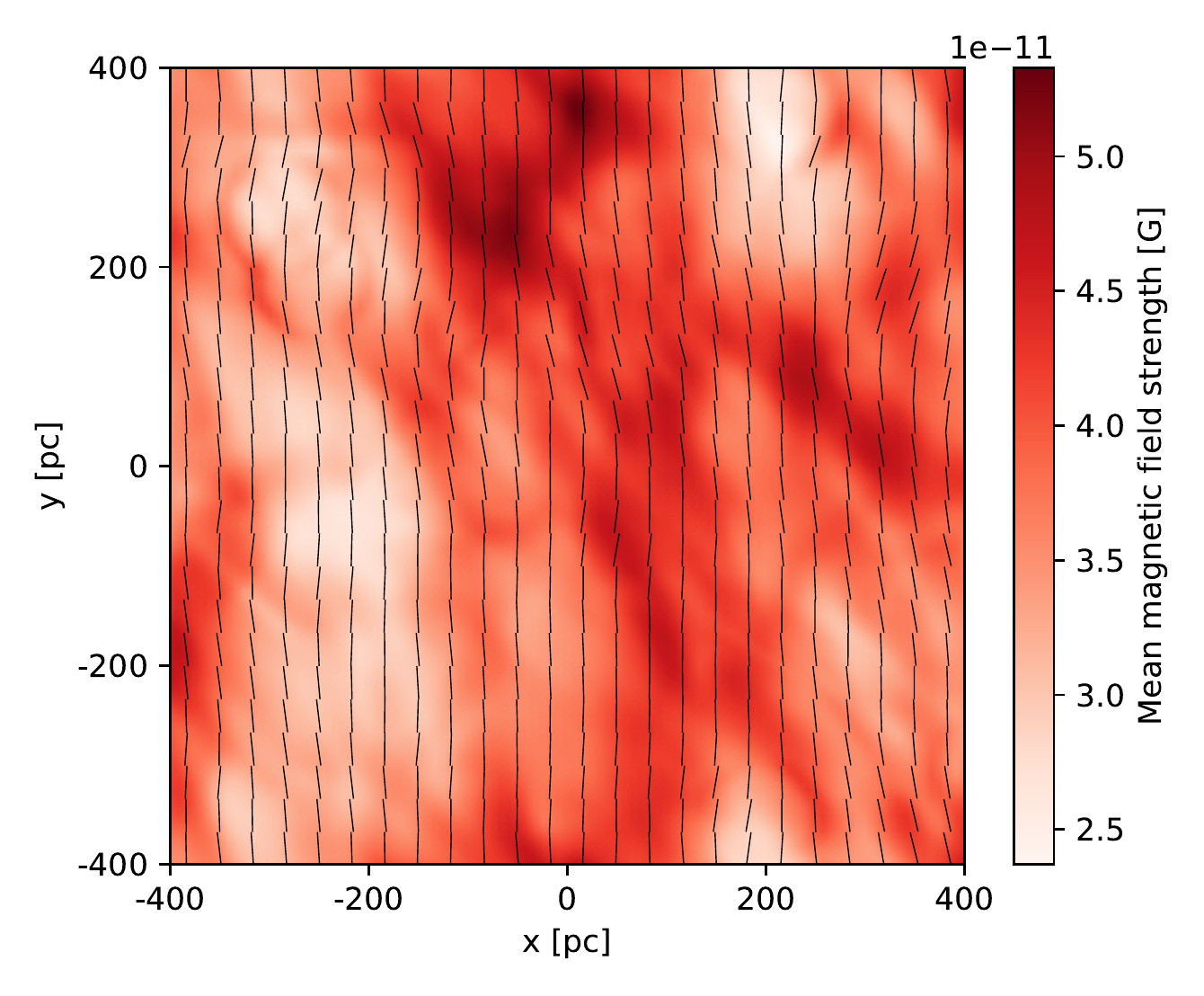}
    \caption{Total magnetic field strength averaged along the $z$ axis for Run25 at $T\simeq 1508$ Myr. 
    Black lines denote the orientation of the projected mean magnetic field in the $x-y$ plane.}
    \label{fig:Mag_proj}
\end{figure}

This setup led to a quasi steady state of kinetic 
and thermal 
energies in all models within first few \Myr, and the 
ISM segregated into multiple phases such as cold dense 
clouds in the midplane, the warm ionised phase, and 
hot ISM bubble in the outer halo of the galaxy. 
Magnetic energy, on the other hand amplified 
exponentially for about a \Gyr, until it reached an 
approximate equipartition with respect to the turbulent
kinetic energy, and thereafter it either saturated or 
kept on amplifying at a drastically slower growth rate
(depending on the SN rate, see \fref{fig:Mag_energy}). 
We refer to these phases as kinematic and dynamical 
phase respectively in the following sections. 
Large-scale magnetic fields of strengths $1-3$\muG and 
of scale-heights $\sim500-
800$\pc were also generated in all models, which 
have previously been analysed and explained as self 
consistent solutions of the mean field dynamo \citep{Bendre2015}. 
In \fref{fig:Mag_energy} we illustrate the two 
different regimes of the dynamo regarding the 
magnetic energy growth, and in \fref{fig:Mag_proj} 
we show the average along the $z$ axis of the 
total magnetic field map in the $x-y$ plane, taken 
in the dynamical phase (at $T\simeq 1508$ Myr) 
for Run25 as a comparison for the observables that 
we will present later.

\section{Observables}
\label{sec_observables}

In this section we summarise the main results of our
analysis, describe how they are obtained, as well as the 
assumptions involved.

\subsection{Rotation measure and Faraday depth}

Any linearly polarised signal can be decomposed into 
two circularly polarised components with opposite 
handedness. Under the effect of a magnetic field, these 
two components accumulate a phase difference and difference in group
velocities 
leading
to a rotation of the plane of 
polarisation.
This is known as Faraday rotation,
and is characterised in the context of ISM by the 
strength of the component of magnetic field parallel 
to the wave vector, wavelength of the radiation, and 
density of scattering thermal electrons.
The projected polarisation angle on the plane of
observation is given by 
\begin{equation}
    \label{theta}
    \theta = \theta_0 + \lambda^2 \mathrm{FD},
\end{equation}
\noindent where $\theta_0$ is the initial angle, 
$\lambda$ the wavelength of the photon and 
$\mathrm{FD}$ is the Faraday depth which
is defined by \citet{burn1966depolarization} as
\begin{equation}
    \label{FD eq}
	\mathrm{FD} = K \int n_e \mathbf{B} \cdot \mathrm{d}\mathbf{l} ,
\end{equation}
where the integration is
along the 
line of sight (LOS) from the source to the observer. 
$K$ 
depends upon the 
natural constants as $K = e^3 / (2\uppi m_e^2 c^4) 
\simeq 0.812$ rad m$^{-2}$ cm$^3$ $\mu$G$^{-1}$ \pc$^{-1}$ 
while
$n_e$ is the thermal electron density. 
The rotation measure ($\mathrm{RM}$) is then 
defined as
\begin{equation}
    \mathrm{RM} \equiv \frac{\mathrm{d} \theta(\lambda)}{\mathrm{d}\lambda^2},
    \label{eq:rm}
\end{equation}
and is equivalent to the FD in the case of a single source along the LOS without 
any internal Faraday rotation and
beam depolarisation.

Throughout this paper we 
only consider LOSs parallel to the 
three common axes of the simulations with observers far in 
the positive direction 
such that every LOS 
along that axis stays parallel to other LOSs along the same axis. 
Furthermore,
in this section
the sources are taken far in the 
negative 
direction.

As the thermal electron density was not explicitly 
computed in the simulations, we model it using the following 
two different prescriptions similar to \citet{RappazEtAL2022}. 
\begin{itemize}
    \item Mod1: 
    the thermal electron density is proportional to the density of the ISM 
    $n_e = c_n n$ 
    with $c_n$\footnote{Note that $c_n$ is constant as long as the mean density of the simulation is preserved. 
    At late times of our simulations a small amount of matter can be lost and thus $c_n$ is adjusted accordingly.} 
    taken such that the mean thermal electron density is $0.1$ cm$^{-3}$ at each time step. 
    \item Mod2:  
    the thermal electron density is taken as constant and set to $n_e = 0.1$ cm$^{-3}$. 
\end{itemize}

\begin{figure}
	\includegraphics[width=\columnwidth]{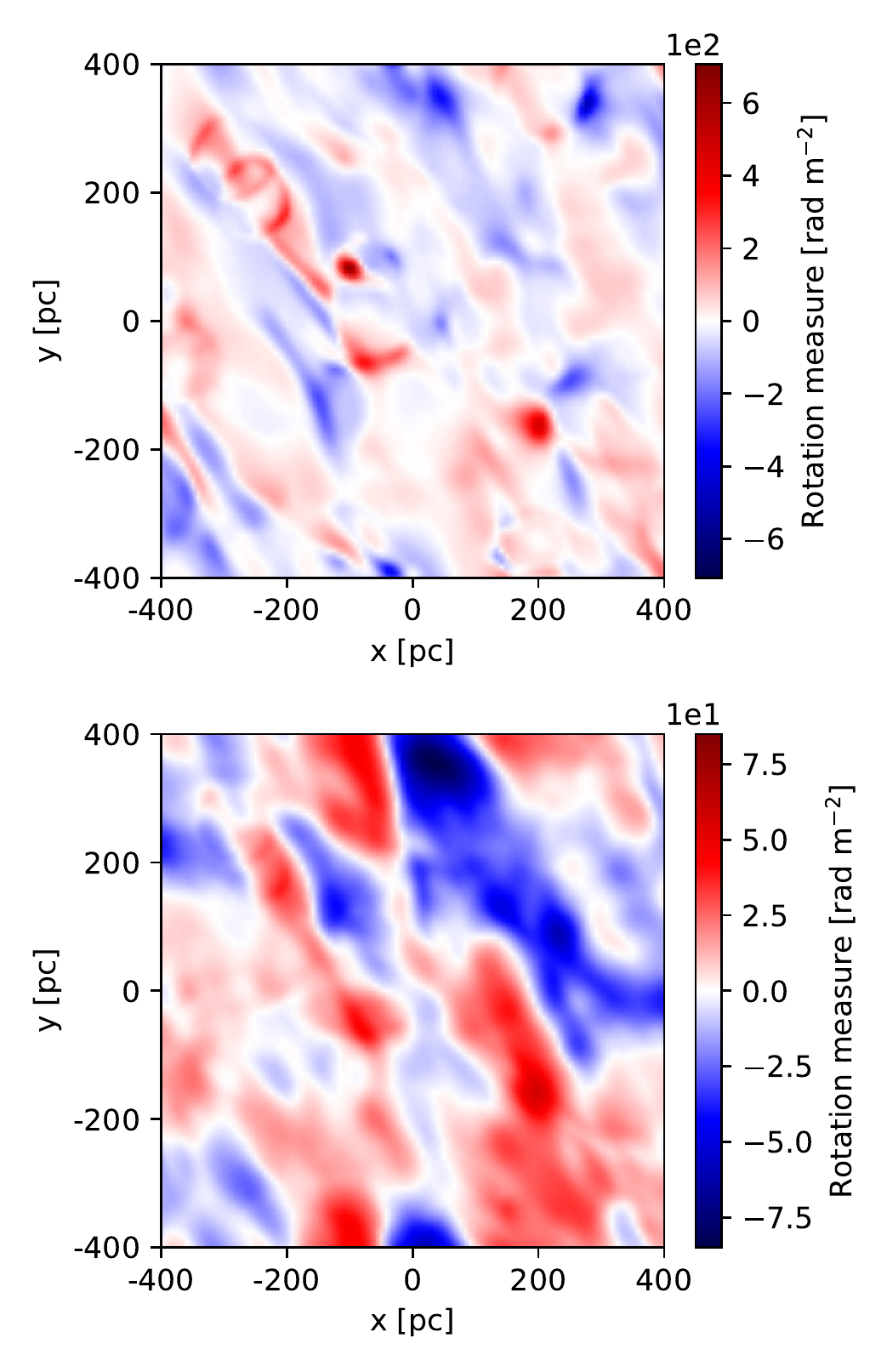}
    \caption{Rotation measure along the $z$ axis for Run25 at 
    $T\simeq 1508$ Myr. This time step is in the middle of the dynamical phase.  
    \textit{Top panel: } Mod1 is used, the thermal electron density is proportional to the ISM density. 
    \textit{Bottom panel: } Mod2 is used, the thermal electron density is constant across the box.}
    \label{fig:RMz with 2 models}
\end{figure}

Using \eref{FD eq}, we then compute and compare the RM maps
obtained with the two aforementioned models of thermal electron densities.
In \fref{fig:RMz with 2 models} we plot the contours 
of $\mathrm{RM}$ along the $z$ axis 
for
Mod1 and Mod2 of $n_e$. 
With Mod2 
the typical values of $\mathrm{RM}$ are almost an 
order of magnitude smaller than that from Mod1, while
the root mean squared (RMS) of the $\mathrm{RM}$ 
with Mod1 is about 
76 rad m$^{-2}$ 
compared to 
20 rad m$^{-2}$ with Mod2. 
With both models of $n_e$ the mean RM is close to 
zero. 
Qualitatively, the observed structures in RM 
maps are larger with the second model as it is sensitive 
only to the magnetic field variations. 
Thus, we expect that the variations of $\mathrm{RM}$ 
in the plane of observation would be important on a larger 
scale
if Mod2 represented the distribution of $n_e$ 
correctly. 
We expect the occurrence of finer structures in the 
RM maps for more complex models of $n_e$, depending upon 
the cross correlations between magnetic fields and $n_e$, 
and their individual distributions as well. 
As such, for Mod1, in which $n_e$ scales
directly with the mass density of hydrogen, we do in 
fact see more small-scale structures (upper panel of 
\fref{fig:RMz with 2 models}). 
In principle,
the scaling between the local distribution of thermal 
electrons and density could also depend on the local phase 
of the ISM.

\begin{figure*}
	\includegraphics[width=2\columnwidth]{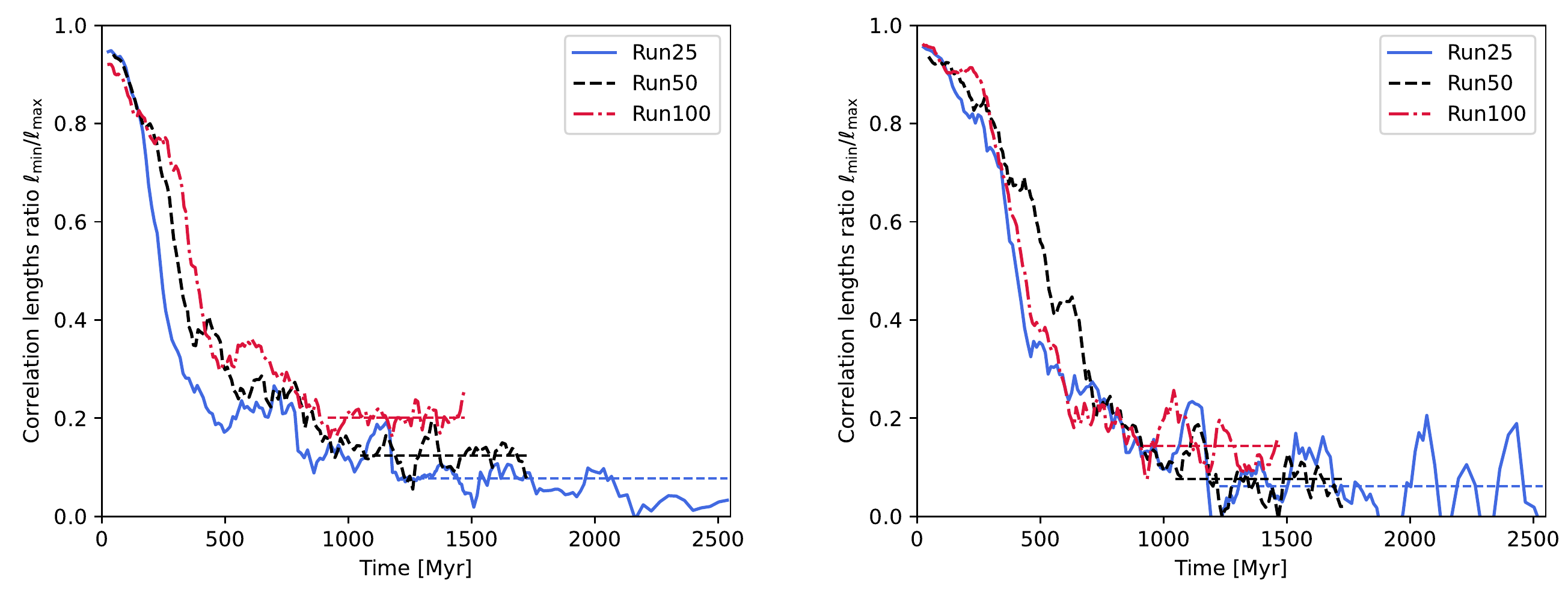}
    \caption{Evolution of the ratio of typical correlation lengths 
    of the rotation measure along the $z$ axis. 
    Each data point is averaged over all data points within 
    50 Myr around its corresponding time. 
    The thin horizontal lines indicate the average value over the dynamical phase. 
    \textit{Left panel:} 
    Mod1 is used to compute the RM.
    \textit{Right panel:} 
    Mod2 is used to compute the RM.}
    \label{fig:RMz_corr_length}
\end{figure*}

A qualitative assessment of the shapes of these structures 
in RM maps suggests that they are 
anisotropic, in a sense that their correlation lengths 
are larger in one direction. This is presumably due to
anisotropy in the magnetic field introduced by
the background shear in the simulations. To analyse this 
assertion more systematically and to quantify the 
anisotropy, we compute the correlation lengths 
$\ell$ from two point correlation functions of RM maps,
which we define for an arbitrary function $f(\mathbf{r})$ as 
\begin{equation}
    C_f(\boldsymbol{R})=\langle f(\boldsymbol{r}) f(\boldsymbol{r}') \rangle = \mathcal{F}^{-1} (|\mathcal{F}(f)|^2), 
\end{equation}
where $\boldsymbol{R}=\boldsymbol{r}-\boldsymbol{r}'$ and $\mathcal{F}$ is the usual Fourier transform. 
This results in elliptical structures centred at $\boldsymbol{r}=\boldsymbol{r}'$ (see e.g. \fref{fig:RMz_corrfct}), which
we normalise and integrate along any particular
line starting from the centre to obtain the correlation 
length in that direction.
We thus obtain the distribution of correlation lengths
as a function of angle with respect to the radial direction.
We summarise in Table~\ref{Table:orientation of RMz} 
the orientation of these ellipses, that is the angle 
between the major-axis of the two point correlation contour with 
$x$ axis in 
degrees. We have computed these angles during the 
kinematic and dynamical phases of the evolution separately, 
along with their $1-\sigma$ variances. 
It does not seem that
model of $n_e$ distribution has any significant impact on 
these orientations in the dynamical phase, and it appears  
that it is entirely decided by the background shear. The angle 
along which the maximum of the correlation lengths is situated, 
matches roughly with the one obtained for background shear 
in \cite{bendre2022}. 
Nevertheless, with increasing SN
rate the direction of the ellipses show larger variations 
of the mean value.
We note however that estimates of variances in Run50 
and Run100 could suffer from a lack of data points, the sampling 
%JS: comma
%being almost half of the one of Run25 and that Run25 is more 
being almost half of the one of Run25, and that Run25 is more 
sampled at the very beginning of the dynamical phase. 
We also show the time evolution of the ratio $\ell_\mathrm{min}
/\ell_\mathrm{max}$ (where  $\ell_\mathrm{min}$ and 
$\ell_\mathrm{max}$ correspond roughly to the minor and 
major axis respectively of the two point correlations of 
RM maps)
in \fref{fig:RMz_corr_length} and
list its averages
over the dynamical phase in Table~\ref{Table:corr_length_ratio} (see also \fref{fig:RMz_corrlength_integral}).

We also study the correlation length of the $\mathrm{RM}$ maps 
in the plane of observation in more detail.
In particular the ratio $\ell_\mathrm{min}/\ell_\mathrm{max}$ shows an interesting behaviour. 
We clearly see in \fref{fig:RMz_corr_length} that for both $n_e$ models in
the early stage of the simulations $\ell_\mathrm{max}$ and $\ell_\mathrm{min}$ are comparable. 
This is mainly due to the initial magnetic field that is uniform 
along the $z$ axis. 
As the system evolves,
the points farther away from each other 
no longer stay
correlated and both $\ell_\mathrm{max}$ and $\ell_\mathrm{min}$ are reduced.
However $\ell_\mathrm{min}$ decreases faster than $\ell_\mathrm{max}$ in the
kinematic phase, indicating that the ellipses are stretched more and
more with time due to differential rotation. 
Once the dynamical phase is reached the ratio is approximately 
constant; mean values of $\ell_\mathrm{min}/\ell_\mathrm{max}$ 
can be found in Table~\ref{Table:corr_length_ratio}. 
These
average values grow non-linearly with the SN rate with a 
scaling slightly steeper for Mod2 than for Mod1. 
Furthermore,
these ratios depend on the model of $n_e$ as well, specifically 
the mean values of $\ell_\mathrm{max}$ are systematically larger 
for the constant $n_e$ model (Mod2), compared to Mod1. This is 
probably owing to the fact that there is an approximate correlation 
between magnetic fields and the mass density ($\mathbf{\it B}\sim 
\rho^{0.5}$), which makes the integrant of \eref{FD eq} proportional 
to $\sim \rho^{1.5}$ in Mod1, while it scales only as $\sim\rho^{0.5}$
for Mod2 (as $n_e$ is constant). 
Regarding the mean values of $\ell_\mathrm{max}$ and 
$\ell_\mathrm{min}$ in the dynamical phase 
(see Table~\ref{Table:corr_length_ratio}), an increase of the SN rate 
reduces the elongation of the RM ellipses. 
The maximum correlation length is particularly affected by the choice of $n_e$. 
We note that $\ell_\mathrm{max}$ compares well with the maximum correlation length of the anisotropic magnetic field reported in \cite{bendre2022}.  
This two-point correlations function, at far
away points from its origin predicts the anti-correlations (note
the negative blue regions in \fref{fig:RMz_corrfct}, outside 
of the central red ellipse, along the $\ell_{min}$), and at some 
time steps these anti-correlations overwhelmingly contribute to 
the integration along $\ell_{min}$, making the correlation length 
negative. Another relevant measure for the correlation length in 
such a case is similar to Taylor microscale, which counts the 
second order moments of the correlation function. It can be 
obtained by fitting a parabola to the curves in 
\fref{fig:RMz_corr_length} and noting where the fit crosses the 
%JS: I think this is better
%$$x$ axis.
$y=0$ line.

Throughout our analysis of the $\mathrm{RM}$ maps we also 
noticed that the $z$ axis is very peculiar.
In fact, it is 
the only axis for which the $\mathrm{RM}$ maps distributions 
are more or less Gaussian
and their standard deviations increase with time and 
are also systematically 
higher in Mod1 than that in Mod2.
The $z$ axis is not directly affected by differential rotation so the random fluctuations of the magnetic field and the thermal electron density along the LOS play a central role in determining the global structure of the map. More importantly these structures are formed on smaller scales than the ones induced by the mean magnetic field and electron density. 

The $\mathrm{RM}$ maps along the $x$ and $y$ axes also 
display elliptical structures elongated along the, 
respectively, $x$ and $y$ axis 
%JS:
%(see \fref{fig:RMx_appendix} and \fref{fig:RMy_appendix}). 
(see Figs.~\ref{fig:RMx_appendix} and \ref{fig:RMy_appendix}). 
Typical structures are 
usually on length scales that are comparable to the 
entire domain for the $y$ axis but not for the $x$ axis (see Table~\ref{Table:corr_length_x_appendix} and Table~\ref{Table:corr_length_y_appendix}). 
These maps are particularly peaked in the galactic 
plane where the magnetic field and density reach their maximum. 
As for the RM along the $z$ axis, we 
also observe some bubble structures that could be due 
to the propagation of matter from SN explosions (shock 
fronts characterised by over dense regions). 
The evolution of the maximal correlation length is 
highly affected by the sign flipping of the $x$ and 
$y$ components of the magnetic field.

\begin{table}
\centering 
\begin{tabular}{c|cccc}
\hline
& & Run25 & Run50 & Run100\\ \hline 
\\[-1em]
\multirow{2}{*}{Kinematic} & Mod1 & $172\pm20$ & $118\pm26$ & $121\pm38$ \\
& Mod2 & $120\pm23$ & $120\pm32$ & $115\pm43$\\\hline 
\multirow{2}{*}{Dynamical} & Mod1 & $119\pm10$ & $125\pm13$ & $126\pm30$ \\
& Mod2 & $120\pm14$ & $123\pm18$ & $125\pm30$\\ 
\end{tabular}
\caption{Comparison for $\mathrm{RM}_z$ of the time average 
of the orientation of the maximum correlation length in 
degree for the two models of the free electron density and the three simulation data sets. 
The errors are taken as one standard deviation of the distribution.} 
\label{Table:orientation of RMz} 
\end{table}

\begin{table}
\centering    
\begin{tabular}{c|cccc}
\hline
& & Run25 & Run50 & Run100\\ \hline 
\\[-1em]
\multirow{2}{*}{$\ell_{\mathrm{min}}/\ell_{\mathrm{max}}$} & Mod1 & $0.077\pm0.020$ & $0.124\pm0.027$ & $0.200\pm 0.018$ \\
& Mod2 & $0.062\pm0.42$ & $0.077\pm0.045$ & $0.144\pm0.044$\\\hline 
\multirow{2}{*}{$\ell_{\mathrm{min}}$ [pc]} & Mod1 & $6.8\pm1.7$ & $10.2\pm2.3$ & $14.8\pm 1.5$ \\
& Mod2 & $6.7\pm5.8$ & $8.2\pm5.1$ & $15.9\pm5.4$\\ \hline
\multirow{2}{*}{$\ell_{\mathrm{max}}$ [pc]} & Mod1 & $97\pm8$ & $92\pm10$ & $80\pm 3$ \\
& Mod2 & $152\pm14$ & $145\pm14$ & $131\pm10$\\ 

\end{tabular}
\caption{Comparison for $\mathrm{RM}_z$ of the
time average in the dynamical phase 
of the correlation lengths ratio 
($\ell_\mathrm{min}/\ell_\mathrm{max}$), the minimum correlation length ($\ell_\mathrm{min}$) and the maximum one ($\ell_\mathrm{max}$)
for the two models and the three simulation data sets.
The errors are taken as one standard deviation of the distribution.} 
\label{Table:corr_length_ratio}   
\end{table}

\subsection{Synchrotron radiation}
\label{synchr theory}

Relativistic charged particles emit synchrotron radiation 
when being accelerated. In the magnetised ISM, they are subject to the 
Lorentz force, so the acceleration is related to the mass and 
the velocity of the particle, as well as to the local magnetic field. 
Most of the synchrotron emission is caused by CR electrons, 
as they are relatively abundant, fast, and light. 
In this analysis we
only consider these CR electrons as a source of the synchrotron 
radiation from our simulation box. 
In the most general case, 
an emitted photon is elliptically polarised in the plane of 
observation and for a group of highly relativistic 
electrons it can be shown 
\cite[see e.g.][]{Westfold1959}
that their total resulting emission is linearly polarised in 
the plane of observation, with a plane of polarisation 
perpendicular to the projected magnetic field. 
The intrinsic 
polarisation angle of the synchrotron radiation is given by 
\begin{equation}
    \theta_\mathrm{i} = \frac{\uppi}{2} + \arctan{\frac{\mathrm{B}_2}{\mathrm{B}_1}},
\end{equation}
where the $\mathrm{B}_{1/2}$ denote the 
magnetic field components along two orthogonal axes 
perpendicular to the LOS. 

As these photons pass 
through the ISM, they also undergo Faraday rotation; 
see \eref{theta}.
The polarisation angle of a cell of synchrotron radiation 
emission is given by 
\begin{equation}
\label{eq:total_ang}
    \theta = \theta_\mathrm{i}  + \lambda^2 \mathrm{FD'}.
\end{equation}
Note that the cell of emission suffers from its own Faraday rotation. 
\citet{Sokoloff1998} showed that in this case we need to subtract 
from the total $\mathrm{FD}$ one half of the $\mathrm{FD}$ 
of the emission cell, we call this quantity $\mathrm{FD'}$ to 
differentiate from Faraday rotation of a non emitting cell. 
Due to the ambiguity by a $\uppi$ rotation of the linearly 
polarised synchrotron emission angle, it is impossible to distinguish 
between coherent magnetic fields, with a constant direction 
along the LOS, and anisotropic magnetic fields, which reverse 
their sign. 

We assume that the CR electron energy spectrum follows
a power law of the form $\mathrm{d}N(E) 
\propto E^{-\gamma} \mathrm{d}E$. Under such assumptions, 
the intrinsic fractional 
polarisation (which does not include any depolarisation 
effect) can be expressed as \citep{LongairBook}
\begin{equation}
    p_\mathrm{i} = \frac{\alpha - 1}{\alpha - \frac{5}{3}},
    \label{he}
\end{equation}
where $\alpha = (1-\gamma)/2$
is the spectral index of the synchrotron radiation. In this paper we consider $\gamma = 2.7$ \citep{Kotera2011}, which gives $p_\mathrm{i} \simeq 0.74$. 

Following an approach similar to \citet{Sur2021} and
\citet{Basu2019} we define the total synchrotron intensity 
map as 
\begin{equation}
    \label{eq synchr int}
    I_{\nu} = \int N_0  n_\mathrm{CR} \mathrm{B}_{\perp}^{1-\alpha} \nu^{\alpha} ~\mathrm{d}l,
\end{equation}
with 
the integration being performed along the LOS, $n_\mathrm{CR} 
= \int N(E)~\mathrm{d}E / \delta^3$, which stands for the 
density of CR electrons, $N_0$ is a
proportionality constant in the CR power law
while $\mathrm{B}_{\perp}$ is the component of 
magnetic field perpendicular to the LOS. 
Note that along the $z$ axis we compute the RM 
through $\mathrm{B}_z$ which is expected to be dominated
by random variations in our set-up. However, the synchrotron emission
is obtained through $\mathrm{B}_x$ and $\mathrm{B}_y$ which capture the
effects of the mean-field dynamo. As such, any quantity derived from
this emission can also be used to characterise the effect of 
the mean-field over the system.

Since the simulations 
did not include CR, similarly to the density of thermal 
electrons, we prescribe the CR electron distribution also
with 
two different models namely:
\begin{itemize}
    \item CRMod1: Here, the CR 
    electron density is proportional to the magnetic energy density. 
    This model assumes that equipartition of 
    CR and magnetic energy is achieved at all scales, and 
    is preserved throughout the evolution.
    \item CRMod2: This model is based on a constant CR electron density.
    The assumption involved here is
    that the CR density variations are either too small 
    or occur on the length scales that are too large to 
    be relevant. 
\end{itemize}

\begin{figure}
\includegraphics[width=\columnwidth]{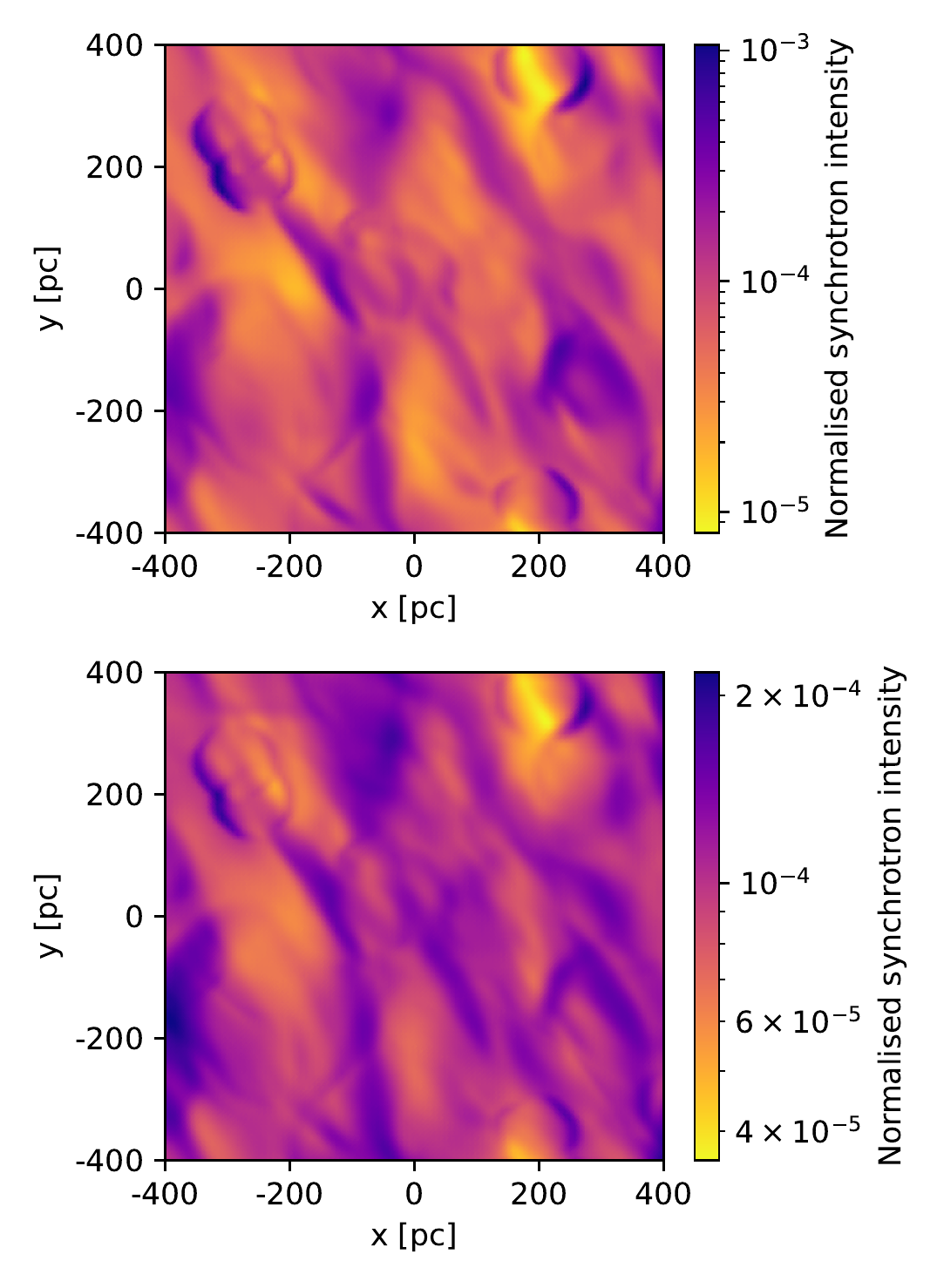}
    \caption{
    Normalised synchrotron radiation intensity along 
    the $z$ axis for Run25 at $T\simeq 1508$ Myr. This 
    time is in the middle of the dynamical phase. 
    \textit{Top panel: } CRMod1 is used, which is based 
    on equipartition between CR and magnetic energy. 
    \textit{Bottom panel: } CRMod2 is used, in which 
    the CR electrons density is constant across the 
    box.}
    \label{fig:Iz_maps}
\end{figure}

\begin{figure*}
	\includegraphics[width=2\columnwidth]{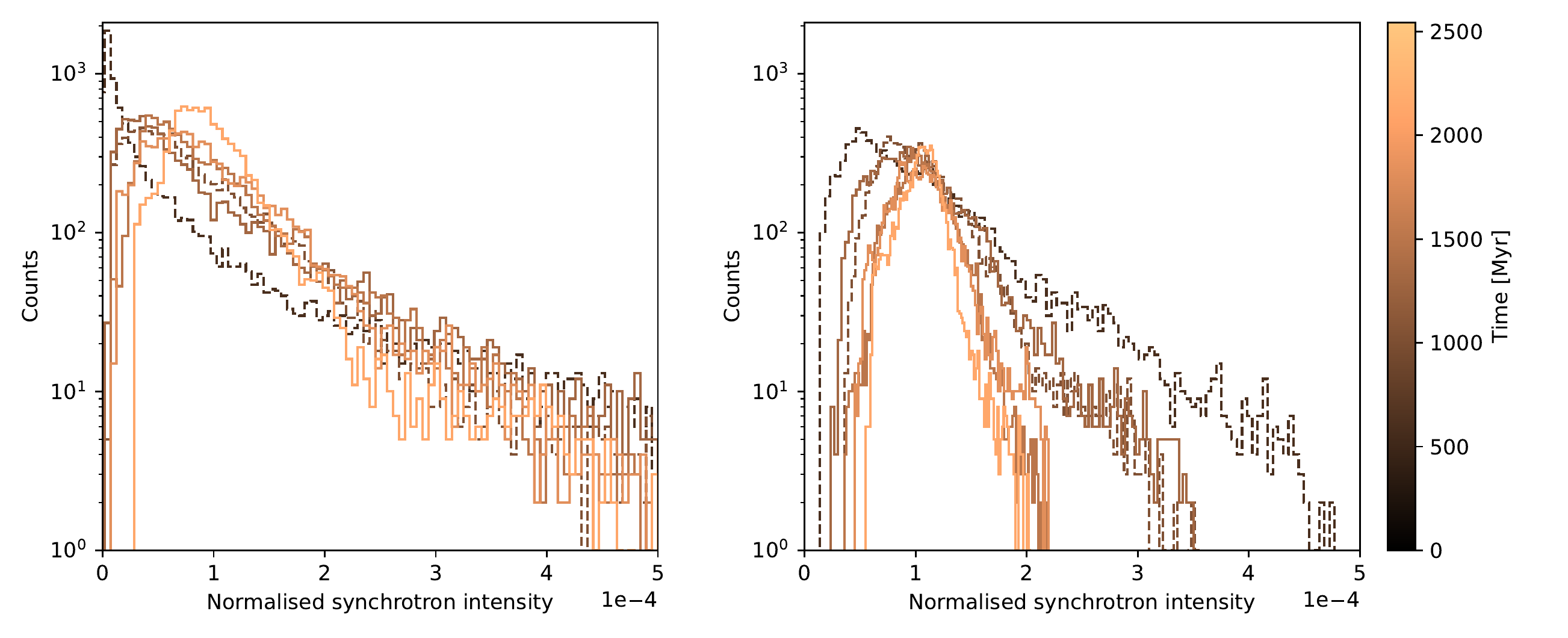}
    \caption{Distributions of the synchrotron radiation 
    along the $z$ axis for Run25 at the frequency $\nu = 1.4$ GHz.
    Histograms shown with dashed lines are taken in the kinematic phase, 
    whereas solid lines correspond to the dynamical phase 
    of the dynamo. 
    \textit{Left panel: } CRMod1 is used, which is based 
    on equipartition between CR and magnetic energy. 
    \textit{Right panel: } CRMod2 is used, in which 
    the CR electrons density is constant across the 
    box.}
    \label{fig:Iz distrib}
\end{figure*}

In order to generalise the discussion we normalise the synchrotron intensity map by its total flux and by this obtain a result that is independent on the frequency and on the CR normalisation $N_0$.
The maps of synchrotron intensity show similar structures 
(see \fref{fig:Iz_maps}, \fref{fig:Ix_appendix} and \fref{fig:Iy_appendix}) with both models, 
although CRMod1 tends to produce a larger range of values 
in synchrotron intensity. 
We illustrate this in \fref{fig:Iz distrib} where we
plot the probability distributions of total synchrotron intensity 
maps integrated along the $z$ axis for both CR electron models, at 
various times. CRMod2 results in narrower 
distributions compared to CRMod1, which could also be ascribed 
to the cross-correlations between the density and the magnetic field inherent for CRMod1.
The synchrotron intensity can be directly
related to the magnetic field for both models this time. 
With 
the models used here
$I_{\nu} \propto \int \left(\mathrm{B}_{\perp}^{3-\alpha} + 
\mathrm{B}_{\perp}^{1-\alpha} \mathrm{B}_{\parallel}^2\right)~\mathrm{d}l$ for CRMod1 and 
$I_{\nu} \propto \int \mathrm{B}_{\perp}^{1-\alpha} ~\mathrm{d}l$ 
in the case of CRMod2. 
So the structures in the synchrotron map trace the
magnetic field lines, that can be stretched due to the 
shearing and randomly twisted by the SN explosions. 
From 
the very similar qualitative aspects of the two maps 
with the two different models of CRs, we conclude 
that it is mainly the magnetic field
strength 
that characterises
the structures. 
In fact, most of the structures in the 
synchrotron intensity maps are also present in the mean 
magnetic field strength over $x-y$ plane 
(see \fref{fig:Mag_proj}).

The $z$ axis is also peculiar regarding the distributions of the 
synchrotron radiation (see \fref{fig:Iz distrib}) as a lognormal 
distribution of synchrotron intensity is observed only for this axis.
It is clear from this plot that the distributions of $I_\nu$ 
starts 
out with a peak at the lower values and with a flat 
tail, and tends eventually to an approximately symmetric shape.
The mean value shifts with time as the magnetic field energy 
is growing. 
Contrary to the $\mathrm{RM}$, the synchrotron radiation 
associated with both models of CR electron distribution is 
not a linear function of 
the magnetic field. 
Therefore, any random fluctuations therein
would not lead to a Gaussian but rather to a lognormal distribution.

We also note that when we  
take the telescope effects (\sref{tel_eff}) into account,
we observe a reduction of the distribution tails, 
since this amounts to filtering over
the small scale structures. 
The inclusion of telescope effects is also accompanied  
by a slight shift of the distributions to the lower normalised synchrotron intensities.

\subsection{Stokes parameters}

The Stokes parameters allow to characterise the polarisation 
state of a beam 
\citep[e.g.][]{taylor2003canadian,haverkorn2006southern,ade2015planck,clark2019mapping}. 
In this paper we 
consider synchrotron emission as the only 
source of the observed radio beam 
and its total intensity is 
given by $I_{\nu}$ as defined in \eref{eq synchr int}. 
The $Q$ parameter is the relative degree of linear 
polarisation along two arbitrary orthogonal axes 
and the $U$ parameter is the relative degree of linear 
polarisation along the same two 
axes rotated by $\uppi / 4$. 
Here, we ignore the
Stokes $V$ parameter
as we are only considering a linearly polarised 
signal. Since we are free 
to choose the orthogonal axes, 
we can conveniently define 
\begin{align}
    Q_{\nu} &= \int p_\mathrm{i} N_0 n_\mathrm{CR} \mathrm{B}_{\perp}^{1-\alpha} \nu^{\alpha} \cos(2\theta) ~\mathrm{d}l,  \\
    U_{\nu} &=    \int p_\mathrm{i} N_0 n_\mathrm{CR} \mathrm{B}_{\perp}^{1-\alpha} \nu^{\alpha} \sin(2\theta) ~\mathrm{d}l,
  \end{align}
with $\theta$ given in \eref{eq:total_ang}.
The polarisation state of the complete beam is affected 
by Faraday effects, 
and the properties of its polarised 
component such as the intensity of the linearly polarised signal 
and the angle of polarisation are defined in terms of the 
Stokes parameters as \citep{gardner1966polarization} 
\begin{align}
\centering
\mathrm{PI}_{\nu} = \sqrt{Q_{\nu}^2 + U_{\nu}^2}, && \Gamma_\nu = \frac{1}{2} \arctan \left( \frac{U_{\nu}}{Q_{\nu}} \right).
\end{align}
We note that, in practice, there is an ambiguity in 
determining this angle since with the ratio $U_{\nu}/
Q_{\nu}$ we lose the information of the 
signs of $U_{\nu}$ and $Q_{\nu}$ separately. 
We thus 
use the function $\textit{arctan2}$ and the modulus of 
$\uppi$ to compute the polarisation angle such that $\Gamma_\nu \in [0,\uppi]$. 
The degree of linear polarisation (DOP) is then 
simply given by $p_\nu = \mathrm{PI}_{\nu}/I_{\nu}$ 
and is in
general decreasing with wavelength due to Faraday effects \citep[see][]{Sokoloff1998}. 

\begin{figure}
	\includegraphics[width=\columnwidth]{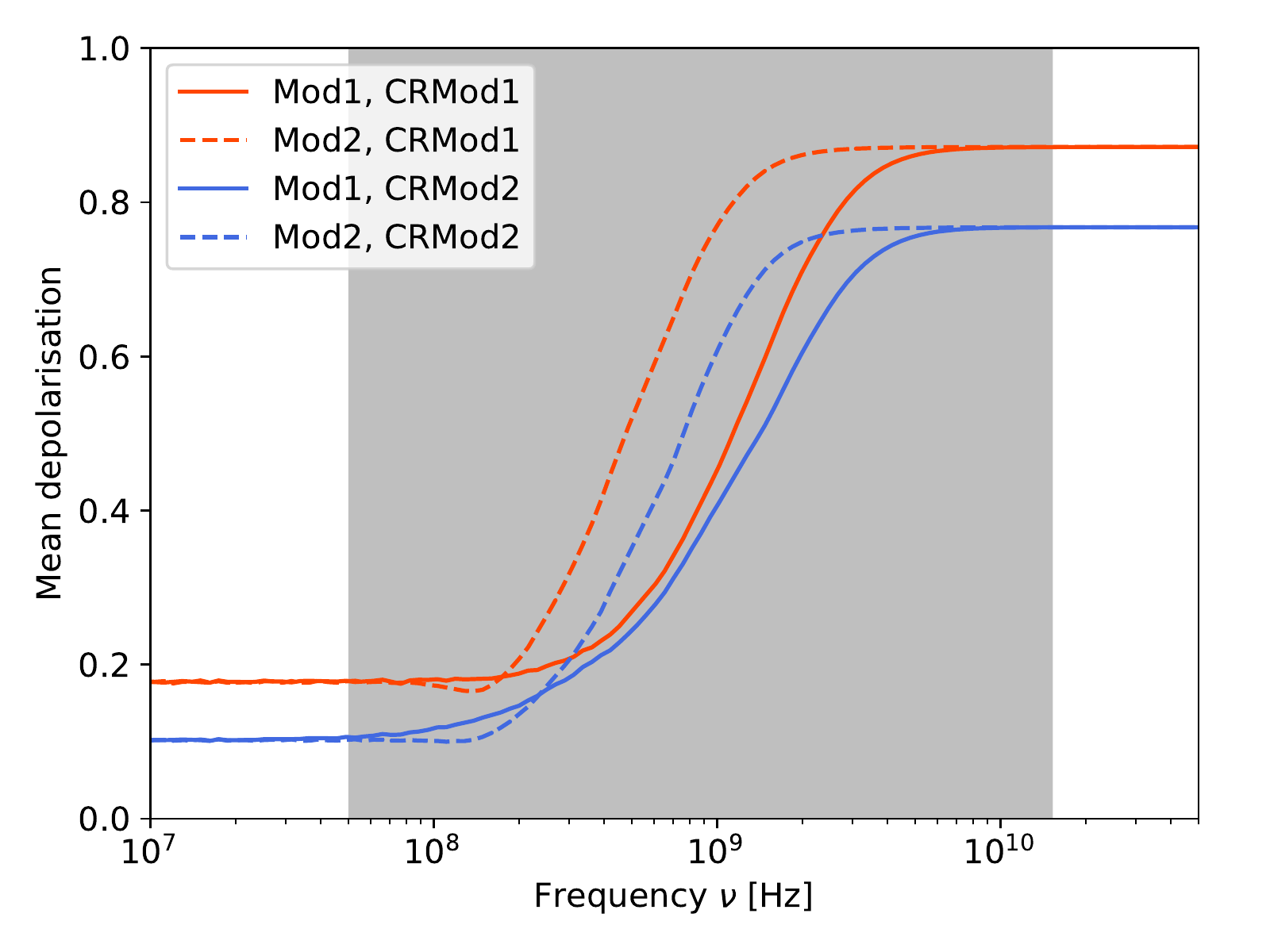}
    \caption{Dependence of the mean observed depolarisation $\langle p_\nu/p_\mathrm{i} \rangle$
    along the $z$ axis on the frequency of observation for the four different configurations of previously defined models for 
    Run25 at $T\simeq 1508$ Myr. 
    The greyed out area represents the frequency range of 
    observation of SKA \citep{dewdney2013ska1}.}
    \label{fig:mean_depo_frequency}
\end{figure}

In \fref{fig:mean_depo_frequency} we show the depolarisation 
which is
the ratio of the
observed DOP and the intrinsic polarisation of synchrotron 
radiation $p_\nu/p_\mathrm{i}$. 
We observe
that in the limits of large and small frequencies it reaches 
constant values
that depend on the CR ray model. 
The depolarisation 
we calculate
here results only from the differential rotation of the plane of polarisation
induced by the Faraday effects and the addition of cells of 
emission along the LOS. 
When the observation frequency 
is very large the Faraday effects are very small and thus 
the depolarisation corresponds to the intrinsic state of 
the magnetic field along the LOS. The value should depend 
upon the distribution of synchrotron intensity and its 
intrinsic angle of polarisation along the LOS. In the very 
low frequency regime the Faraday effects are so large 
that any correlation between the different intrinsic 
angle of polarisation is lost, which leads to a large 
decrease in the DOP. Its value then tends to the one obtained with a 
completely random distribution of angles. In the 
intermediate regime however, the evolution of the 
mean depolarisation with the frequency is mainly 
dependent on the thermal electron distribution model. 
Since we have a certain distribution of FD along
the LOS, the frequency at which the term $\mathrm{FD} 
\lambda^2$ becomes negligible is not the same for each 
cell of synchrotron emission. 
Especially, the frequency at which the mean depolarisation starts
to increase, and thus the transit from one constant regime to the other
(see \fref{fig:mean_depo_frequency}), will give the order of magnitude 
of the lowest absolute values of FD encountered 
along the LOS in the plane where the average is performed. 
The same is also true for the upper bound
of the transition but this time for the highest values of FD.

\begin{figure*}
	\includegraphics[width=2\columnwidth]{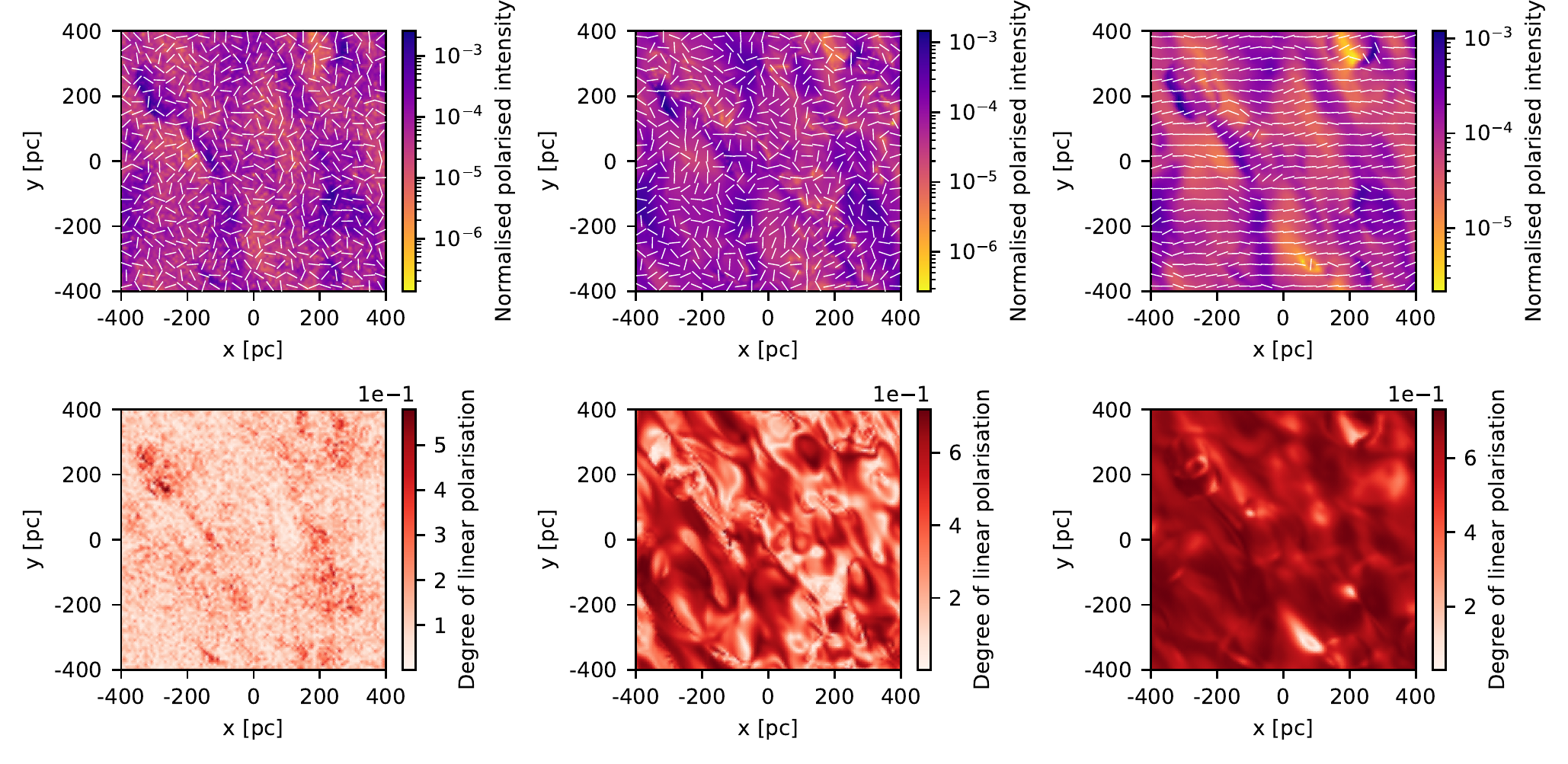}
    \caption{\textit{First row:} Normalised polarised intensity maps $\mathrm{PI}_{\nu}$ 
    with respect to the $z$ axis. The lines indicate the observed angle of polarisation $\Gamma_\nu$ with respect to the $z$ axis. \textit{Second row:} Degree of linear polarisation $p_\nu$ with respect to the $z$ axis. 
    The three columns correspond, respectively, to a frequency of observation of 
    70~MHz, 1.4~GHz, and 5~GHz for Run25 at 
    $T\simeq1508$ Myr, i.e.~in the dynamical phase. 
    We used Mod1 and CRMod1 to compute all the presented quantities.}
    \label{fig:maps_1}
\end{figure*}

We find that the typical structures in the total synchrotron 
intensity maps are very similar to the ones observed for the 
polarised intensity for very high frequency. In the first 
row of \fref{fig:maps_1} we show the observed polarised 
intensity along the $z$ axis at, respectively, 
70~MHz, 1.4~GHz and 5~GHz
using the configuration of models that 
is perhaps more realistic (Mod1 and CRMod1). 
We observe throughout these typical maps that the nature of 
observed structures depends strongly on the frequency of 
observation. Especially on low frequencies we completely 
lose the details on small spatial scales. 
We also note 
that our normalisation of the total incoming flux due to the 
synchrotron effects is arbitrarily fixed, so the typical 
values that we obtained should not be directly compared to 
real observations, which is why we normalise each map by its total flux. 
The spacial distribution of these structures, however, are
unaffected by this normalisation. 

The DOP and the observed polarisation angle are also affected by 
the frequency 
(row 2 and lines in row 1 of \fref{fig:maps_1}). Contrary to 
the polarised intensity maps (in row 1), the structures are 
no longer smoothed when the frequency is decreased but are
mostly destroyed. 
In the very low frequency regime, the typical 
structures are of the order of a pixel, which we 
also confirm from
the evolution of the correlation length as a function of the 
frequency of observation. 
There, a similar transition to the 
one presented in \fref{fig:mean_depo_frequency}
is observed. 
Furthermore, the DOP maps in the intermediate frequency 
regime show new elongated structures that are oriented 
with respect to the radial ($x$) axis, which could be due  
to the differential rotation.

In the high frequency regime the DOP is relatively uniform 
due mainly to the correlation of the orientation of the 
projected magnetic field along the LOS. 
In particular, at 5~GHz the observed angle of polarisation
is almost always perpendicular to the orientation of the 
projected mean magnetic field (see \fref{fig:Mag_proj}) since the Faraday
effects are small. However, when the frequency of observation is decreased
this statement no longer holds and the correlation between the polarisation
state and the magnetic field is lost.
In Appendix \sref{Sec:appendix_obs} we also display the equivalent to \fref{fig:maps_1} for the $x$ and $y$ axes (\fref{fig:few_maps_appendix_x} and \fref{fig:few_maps_appendix_y}).

\section{Telescope effects}
\label{tel_eff}

In this section we 
describe the effects of a finite telescope resolution on 
the observables extracted from the simulations. 
In order to make a comparison with actual observations we smooth 
the observed maps with a Gaussian kernel \citep{telescopeBook},
with the aim of simulating the aforementioned effect of 
finite telescope resolution.
We use a convolution kernel 
with a full width at half-maximum (FWHM) being the same in all 
directions. 
It mainly leads to a deletion of the small scale 
structures while preserving the larger ones.
However, we need to treat the DOP and polarisation angle maps 
more carefully as they are obtained through the transformation 
of the actual observables, i.e. the Stokes parameters $Q$ and 
$U$. In practice, observations 
thus depend upon the frequency of observation as well as the 
resolution of the used telescope.

\begin{figure}
	\includegraphics[width=\columnwidth]{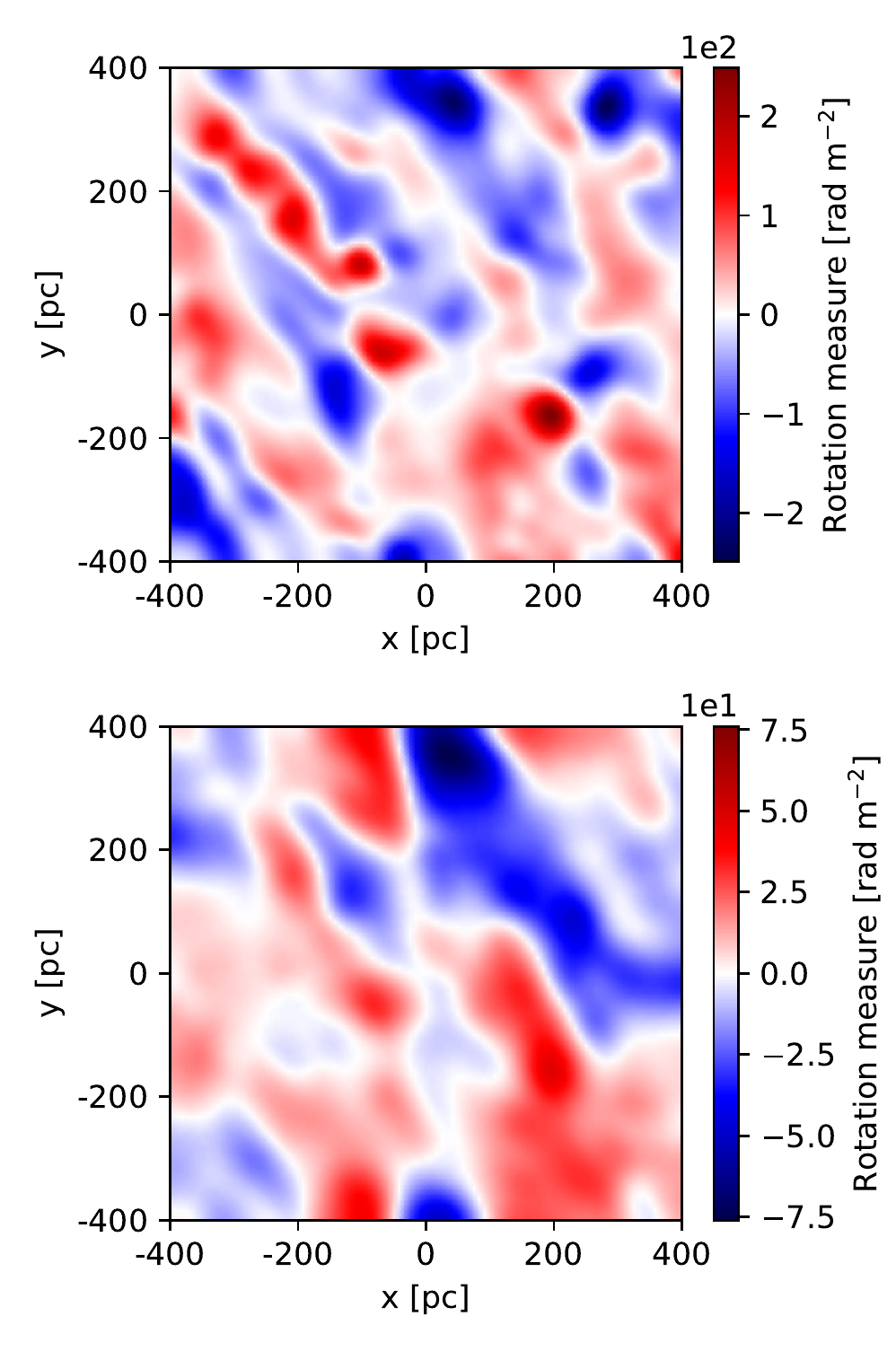}
	
    \caption{Rotation measure along the $z$ axis for Run25 at 
    $T\simeq 1508$ Myr. This time step is in the middle of the dynamical phase. The two maps are smoothed by a Gaussian kernel of $5\times 5$ pixels.
    \textit{Top panel: } Mod1 is used, the thermal electron density is proportional to the ISM density. \textit{Bottom panel: } Mod2 is used, the thermal electron density is constant across the box.}
    
    \label{fig:RMz with 2 models telescope}
\end{figure}

We set the FWHM of the Gaussian kernel to 5 pixels which 
corresponds to a smoothing scale of $\sim 42$\pc. If we 
consider, for example, that the telescope can achieve a 
resolution of 0.5 arcsec, 
the resolution of $\sim42
$\pc would locate the galaxy
at $\sim 17$ Mpc. 
When comparing the properties
of RM maps with (\fref{fig:RMz with 2 models telescope}) 
and without (\fref{fig:RMz with 2 models}) the telescope 
effects we see that we have lost the very small scale 
details like the bubbles structures in the upper left 
corner of Mod1. The RMS values are reduced to 
57
and 
18
rad m$^{-2}$ for Mod1 and Mod2, respectively. 
On the other hand, we recover qualitatively the overall 
information encoded in the large scale ellipses of 
the maps, like the sign of the mean magnetic field 
along the LOS or the orientation of the structures. Note 
however that the peaked areas of the RM 
map associated with 
Mod1 have been erased. The length scales are not affected 
significantly but are
increased by $\sim 5$ pc
which is about 5\% of its original value. 
We also note that a study of the RM at a significantly 
higher
redshift would require a modification of the \eref{FD eq}
\citep[see e.g.][]{2011ApJ...738..134A}. 

\begin{figure*}
	\includegraphics[width=2\columnwidth]{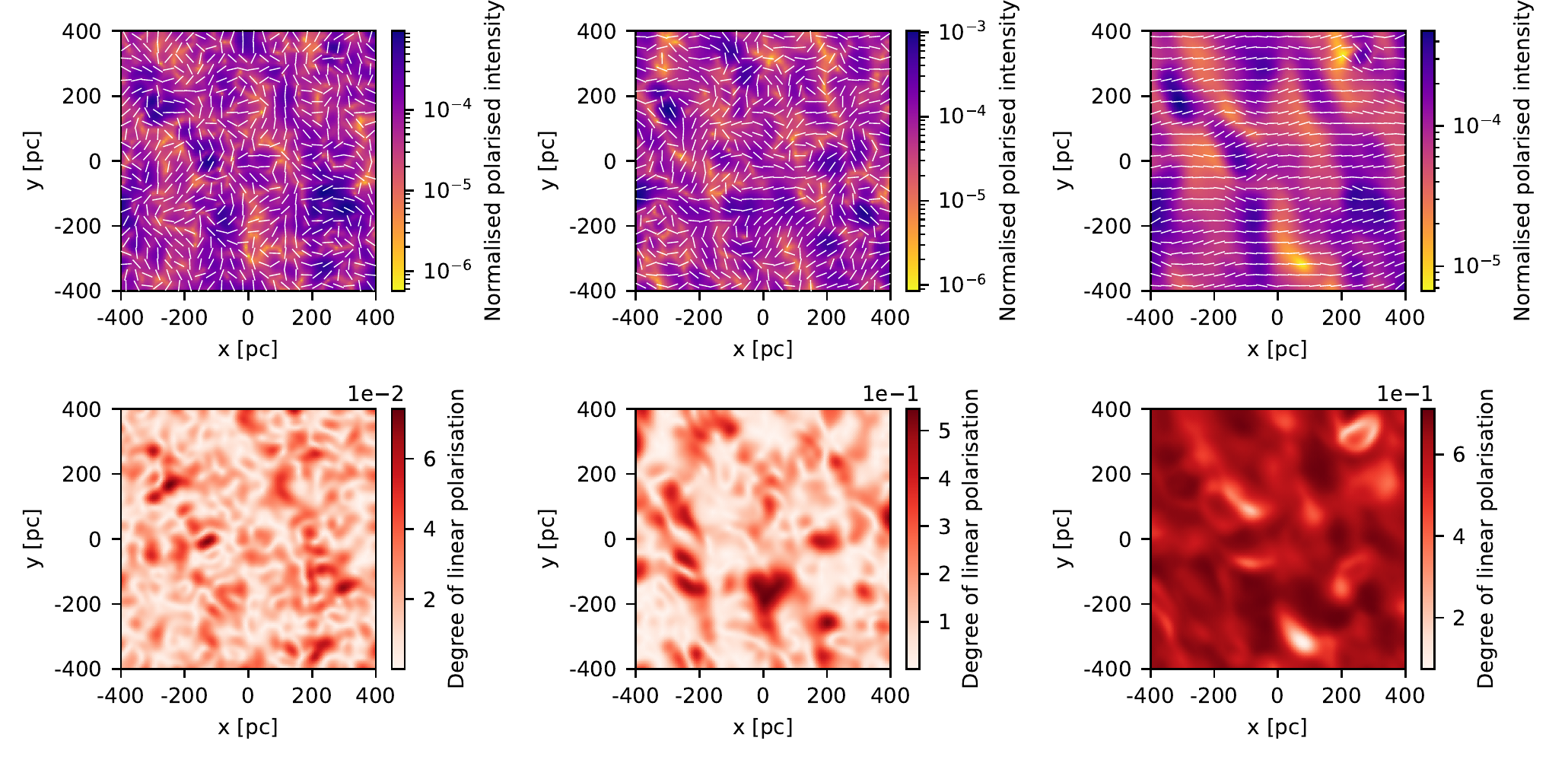}
    \caption{\textit{First row:} Normalised polarised intensity maps $\mathrm{PI}_{\nu}$ 
    with respect to the $z$ axis. The lines indicate the observed angle of polarisation $\Gamma_\nu$ with respect to the $z$ axis. \textit{Second row:} Degree of linear polarisation $p_\nu$ with respect to the $z$ axis. 
    The three columns correspond, respectively, to a frequency of observation of 
    70~MHz, 1.4~GHz, and 5~GHz for Run25 at 
    $T\simeq1508$ Myr, i.e.~in the dynamical phase. 
    We used Mod1 and CRMod1 to compute all the presented quantities. All six maps are obtained after smoothing the observable with a Gaussian kernel with FWHM of 5 pixels.}
    \label{fig:maps_2}
\end{figure*}

In \fref{fig:maps_2} we present the exact same maps as in 
\fref{fig:maps_1} but corrected by telescope effects. We 
clearly see that the filaments and arc structures in 
the polarised intensity contours are still identifiable 
but thickened, thus the small scale details are 
not
distinguishable from the background. 
There are also 
differences in the evolution regarding the frequency of
observation. 
If we focus on the lowest values of the maps 
we see that they are spatially distributed in such a way that
they delimit structures of higher values without agglomerating 
or forming large scale structures themselves. 
It is then even harder to disentangle
between the original peaked structures and the new structures induced by
the Gaussian smoothing.
The effect on the DOP and angle of polarisation are even 
more pronounced as we cannot achieve the same random 
pixel-wise maps on frequencies $\nu \sim 70$~MHz.
In particular, telescope effects lead to the creation 
of structures that we do not observe in maps with the 
maximum resolution of the simulation, i.e.~$\delta 
\sim 8~\mathrm{pc}$. However, it seems that at very 
low frequencies the isotropy observed at the maximum 
resolution is still observed with telescope effects for these two quantities. 
Furthermore, we observe that the typical values of DOP 
and normalised polarised intensity are reduced by 
Gaussian smoothing.

\begin{figure}
	\includegraphics[width=\columnwidth]{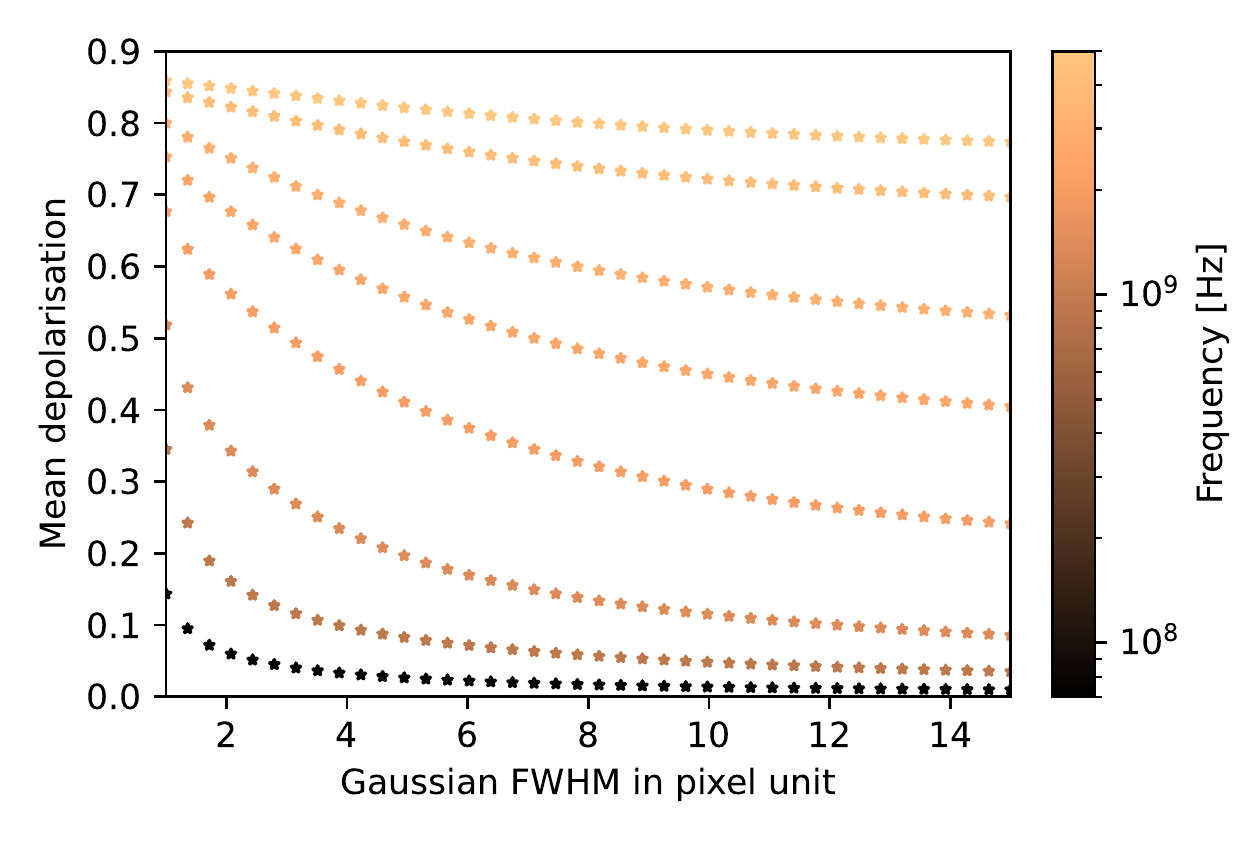}
    \caption{Evolution of the mean depolarisation $\langle p_\nu/p_\mathrm{i} \rangle$
    along the $z$ axis for Run25 at $T\simeq 1508$ Myr as a function of the Gaussian kernel used to simulate telescope effects. The depolarisation is computed using Mod1 and CRMod1.}
    \label{fig:mean_depo_telescope}
\end{figure}

The overall shapes of the curves in \fref{fig:mean_depo_frequency} 
are not affected by the telescope effects; 
indeed the constant value of the mean depolarisation
in the high and low frequency limit still depends 
mainly on the 
CR distribution model, and the intermediate transition region 
still depends on the model of thermal electrons.
However we still can notice differences. 
The mean depolarisation at low frequencies is significantly 
lower and closer to zero. 
The frequency range of transition is smaller, especially the end of the
transition at low frequencies occurs at a higher frequency
($\nu\sim 0.5$~GHz compared to $\nu\sim 0.1$~GHz without telescope effects). 
This trend is also confirmed in \fref{fig:mean_depo_telescope}, 
where we show the evolution of the mean depolarisation as we 
increase the FWHM of the Gaussian kernel and the frequency of observation. 
This evolution is not linear and highly depends on the frequency of observation.
The telescope effects are particularly important 
at low and intermediate observation frequencies, 
as we could directly see from the DOP maps 
(see second row of \fref{fig:maps_2}). 
It also supports the idea that when we have a finite resolution 
observation we lose more the information generated by the 
lowest values of FD along the LOS. This is different from the RM maps where we lost most of the information on the structures generated by the highest values of RM. 

\section{Discussion}
\label{sec_discussion}

In this section we discuss the main limitations of 
this work.
First, we would like to 
point out that the simulations domain is 
much smaller than a typical disc galaxy in $x$ 
and $y$ directions. For example the Milky Way has a 
radial length scales of $\sim 30$\kpc, as opposed to 
$\sim 0.8\,\times\, 0.8$ \kpc spanned by the simulation 
domain in $x$ and $y$ direction.  Any extracted observable thus would therefore 
reveal the local correlations in the ISM turbulence, and 
any correlations exceeding a \kpc scale would not be 
captured. Furthermore, the reconstruction of RM maps 
of an entire galaxy would also need the radial dependence 
of gas distribution and any azimuthal asymmetries to be 
taken into account \citep{gasdistrib}.
That is the main reason we performed our analysis along
the $z$ direction. 

We already mentioned that the simulations analysed only include neutral hydrogen 
but in reality several species could have non negligible contribution 
to the ISM dynamic. 
Especially this treatment does not allow us to get a self consistent 
evaluation of the thermal and CR electrons densities. 
We had to define very simple models for both of them. 
The mean free electron density used here is set to 
0.1 cm$^{-3}$, especially with Mod2 we neglect any
kind of spatial and temporal variations. 
From observations we know that in spiral galaxies the 
mean thermal electron density can range between
$\sim$ 0.1-0.01 cm$^{-3}$ in the disc plane depending 
if we consider arms or inter-arm regions \citep{schnitzler,Yao,beck2019synthesizing}.
However, the free electron density is expected to be even smaller outside the
disc plane. As the scale height of the simulation box is larger than a typical disc,
the mean free electron could be overestimated.
With Mod1 we neglect contributions from other parameters, such as temperature, to 
evaluate the local thermal electron density.

Even stronger assumptions were made regarding the CR models. A constant CR model would hold only on very small scales and not at all on a galactic scale. 
On the contrary, equipartition seems not to hold on scales smaller than $\sim 1$ kpc \citep{equipart1kpc,equipart1kpc2}, whereas in this work our maximum resolution for any quantity is about 8.3 pc. 
We would need a more rigorous model that lies in between our two hypothesises. 
As an example a more complete description is proposed by \citet{SchoberCR} that includes contributions to the CR spectrum from the main interactions of electrons in galaxies. 
However, this model does not include the diffusion of cosmic rays \citep[see e.g.][for an analysis of the CR diffusion]{sampson2022turbulent}.
Treating the CRs directly in the simulations can also affect 
the magnetic field itself \citep{10.1093/mnras/staa3509} which eventually affects all the observables presented. 

We also would like to review effects that can modify the slope of the synchrotron emission spectrum, we used a constant value of $\alpha = -0.85$ which would correspond to a typical value in the GHz frequencies \citep{Platania_1998}. 
In practice when going to frequencies $\nu \ll$ GHz the spectrum is modified due to the contributions from synchrotron self-absorption and free-free absorption. 
In the opposite when going higher than a few GHz we should include effects of free-free emission, eventually at very high frequency thermal emission could also be considered. 
These different interactions contribute together to flatten the synchrotron spectrum \citep{Guzman,kogut2012synchrotron}. 
It could be a straight forward extension of this work to see how the observables and observed structures are affected by these effects at large and low frequencies.

\section{Conclusions}
\label{sec_conclusions}

In this paper we have explored the typical observational 
signatures of a galactic magnetic field that has been 
self-consistently generated by a large-scale turbulent 
dynamo. In particular we looked at three types of 
observables namely the Faraday RM, the synchrotron 
radiation, and the Stokes parameters, as well as their 
related quantities such as the DOP (\sref{sec_observables}). 
In the second part we applied simple telescope effects 
(\sref{tel_eff}) to mimic radio observations. 

The magnetic field was directly obtained from a simulation 
of a galactic dynamo, while other relevant quantities for 
galactic radio emission were not included in the simulation. 
In particular, we had to model the distribution of thermal and 
non-thermal electrons. We have defined two models of thermal 
electrons, one with a 
constant electron density while in the other one we let 
the electron density scale with the local mass density. 
Similarly for CR electrons density, we used a model 
with constant CR density and in the other one we scaled
it with magnetic energy density.

We found that the RM structures are consistent with the 
direction of differential rotation in the $x-y$ plane. 
The SN rate does not seem to have an influence on the 
overall orientation of the structures in the dynamical
phase. 
However it was noticed that the stability and 
the typical sizes of these structures are affected 
by the SN rate. 

Our results also indicate that the RM is sensitive to the 
choice of the thermal electron model both in terms of qualitative
aspect of the structures and typical values of the maps. 
From the RM distribution we found that along an axis not 
directly 
influenced by rotation measure (i.e.~the $z$ axis in our 
setup) it is the magnetic field and thermal electron density 
variations that dominate
the resulting maps.

Synchrotron radiation intensity maps showed the same structures 
as in the mean magnetic field maps.
We found that the final qualitative aspect of the maps almost does not depend 
on the choice of the model of CR electron. 
However distributions 
showed a significantly increased degree 
of symmetry with the constant 
CR model along the $z$ axis. 
Maps of quantities that are derived from the Stokes parameters maps are highly dependent 
on the frequency of observation. We found that decreasing the 
frequency introduces something similar to a background noise 
over the maps, deleting structures. It was also observed that 
at an intermediate frequency completely new structures could 
appear, this frequency depends on the typical FD encountered 
in the photon's path.

Under telescope effects at a resolution of FWHM $\sim 42$ pc, we observed mainly deletion of small scale spatial structures in every 
map. 
We also found that these effects depend non-linearly on the 
frequency of observation. 
In particular when we applied telescope effects, completely new 
structures 
in the contours of various quantities related to the Stokes parameters were observed at low frequencies.
In the low frequency regime, the typical structures are mainly 
induced by FD values for which the term $\lambda^2 \mathrm{FD'}$ 
(with $\mathrm{FD'}$ being the FD for a self-emitting cell) is 
neither
too large nor negligible.
When this term is very large the cell of emission loses its correlation
with the other cells along the LOS, thus a tiny variation in the frequency 
of observation will result in a random modification of Stokes $Q$ and $U$. 
On the other hand when it is small, the cell of emission will not modify Stokes 
$Q$ and $U$. 
Finally, if FD is such that $\lambda^2 \mathrm{FD'}$ is somewhere in-between, 
modifications in Stokes $Q$ and $U$ due to a frequency variation are smooth. 
As such the telescope effects seem stronger at the 
lower frequencies, the Gaussian smoothing tends to alter 
the information given by lowest values of FD more than 
high values. However we observed in RM maps that the 
deletion of small scale details happened primarily on 
the highest values of RM. 

With this work we would like to motivate 
observational analyses that study the RM and Stokes 
parameters simultaneously, as they give complementary 
information about the magnetic field.
We would like to emphasise that at low frequencies 
observations might suffer from the effect of a finite 
resolution and may therefore lead to 
wrong conclusions. 
With the new generation of telescopes, 
such as the Square Kilometre Array (SKA), however, the 
resolution of radio observations will highly increase 
which might
allow lower frequencies to be explored increasing the 
comprehension of galactic magnetic fields.

\section*{Acknowledgements}
%JS: added Yoan
%We are grateful to Kandaswamy Subramanian for providing very
We are grateful to Kandaswamy Subramanian and Yoan Rappaz for providing very
useful comments on our manuscript.
A.B.~and J.S.~acknowledge the support by the Swiss National
Science Foundation under Grant No.\ 185863.

\section*{Data Availability}

%JS: To be checked with Abhijit
% The simulation data are publicly available at \href{http://silcc.mpa-garching.mpg.de}{http://silcc.mpa-garching.mpg.de} under data release 6 (DR6, \citealt{SILCC_project_molecularclouds}). 
%JS: shorter:
% Data used in this analysis will be provided upon reasonable request.
% The analysis scripts for this study will be shared upon 
% request to the corresponding author.
Data used in this analysis scripts will be provided upon reasonable request to the corresponding author
%JS.



\bibliographystyle{mnras}
\bibliography{example} % if your bibtex file is called example.bib

%\begin{thebibliography}{99}
%\end{thebibliography}

%%%%%%%%%%%%%%%%%%%%%%%%%%%%%%%%%%%%%%%%%%%%%%%%%%

%%%%%%%%%%%%%%%%% APPENDICES %%%%%%%%%%%%%%%%%%%%%

\appendix
\section{Two-point correlation of the RM}
\label{Sec:appendix_RMz}
\begin{figure}
	\includegraphics[width=\columnwidth]{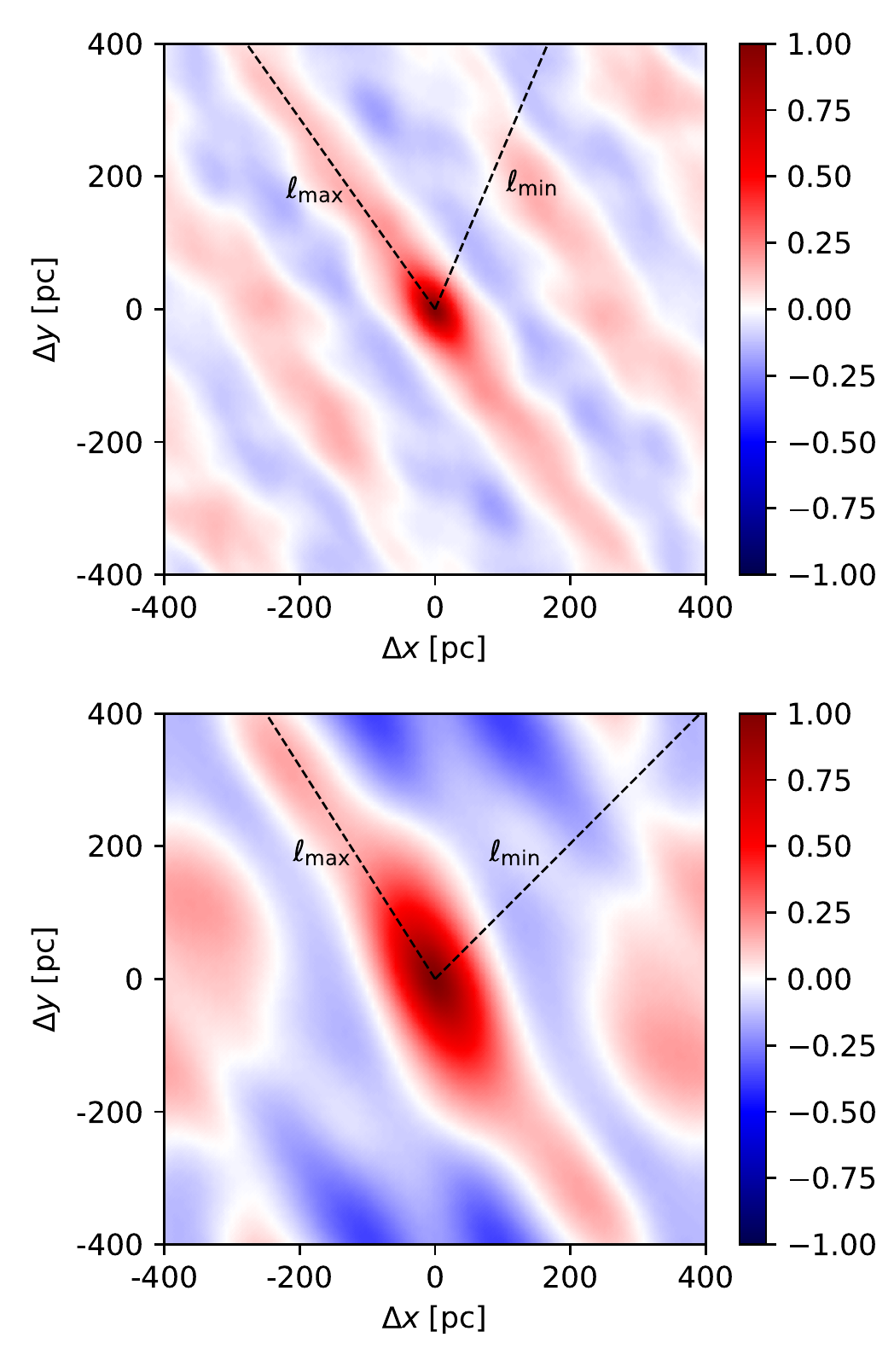}
	
    \caption{Rotation measure correlation function $C_{\mathrm{RM}}$ along the $z$ axis for Run25 at 
    $T\simeq 1508$ Myr. This time step is in the middle of the dynamical phase.
    \textit{Top panel: } Mod1 is used, the thermal electron density is proportional to the ISM density. \textit{Bottom panel: } Mod2 is used, the thermal electron density is constant across the box.}
    
    \label{fig:RMz_corrfct}
\end{figure}

\begin{figure}
	\includegraphics[width=\columnwidth]{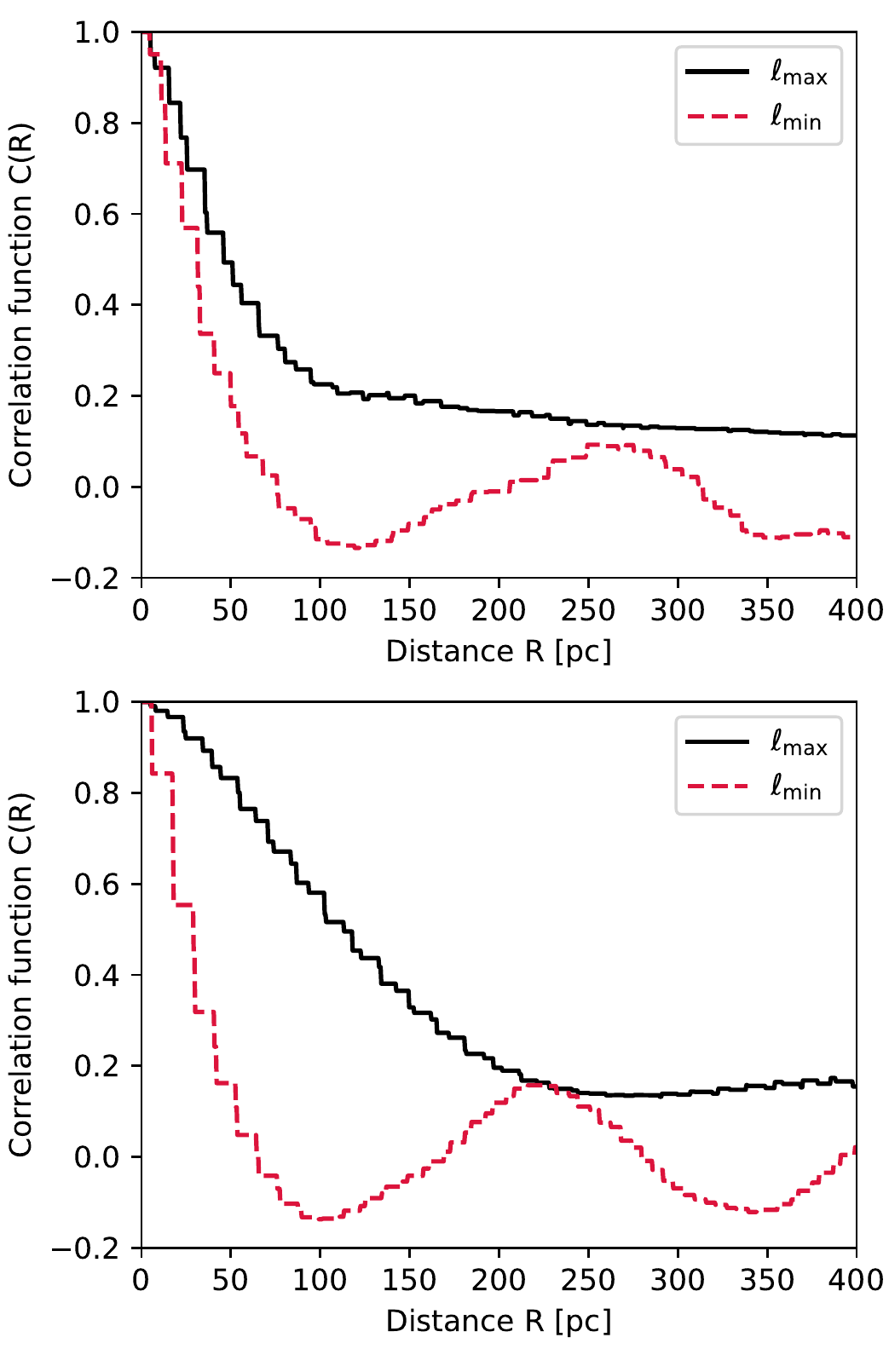}
	
    \caption{Rotation measure correlation function $C_{\mathrm{RM}}$ along the axes of minimal and maximal correlation length displayed in \fref{fig:RMz_corrfct}. The RM is computed along the $z$ axis for Run25 at 
    $T\simeq 1508$ Myr. This time step is in the middle of the dynamical phase.
    \textit{Top panel: } Mod1 is used, the thermal electron density is proportional to the ISM density. \textit{Bottom panel: } Mod2 is used, the thermal electron density is constant across the box.}
    
    \label{fig:RMz_corrlength_integral}
\end{figure}

In this section we illustrate the way we compute the correlation length.
\fref{fig:RMz_corrfct} represents an example of two-point correlation maps for the RM along the $z$ axis.
\fref{fig:RMz_corrlength_integral} shows the correlation function along the axes of maximum and minimum correlation length, which are simply obtained by integration.

\section{Observables along radial and azimuthal directions}
\label{Sec:appendix_obs}

\begin{figure}
	\includegraphics[width=\columnwidth]{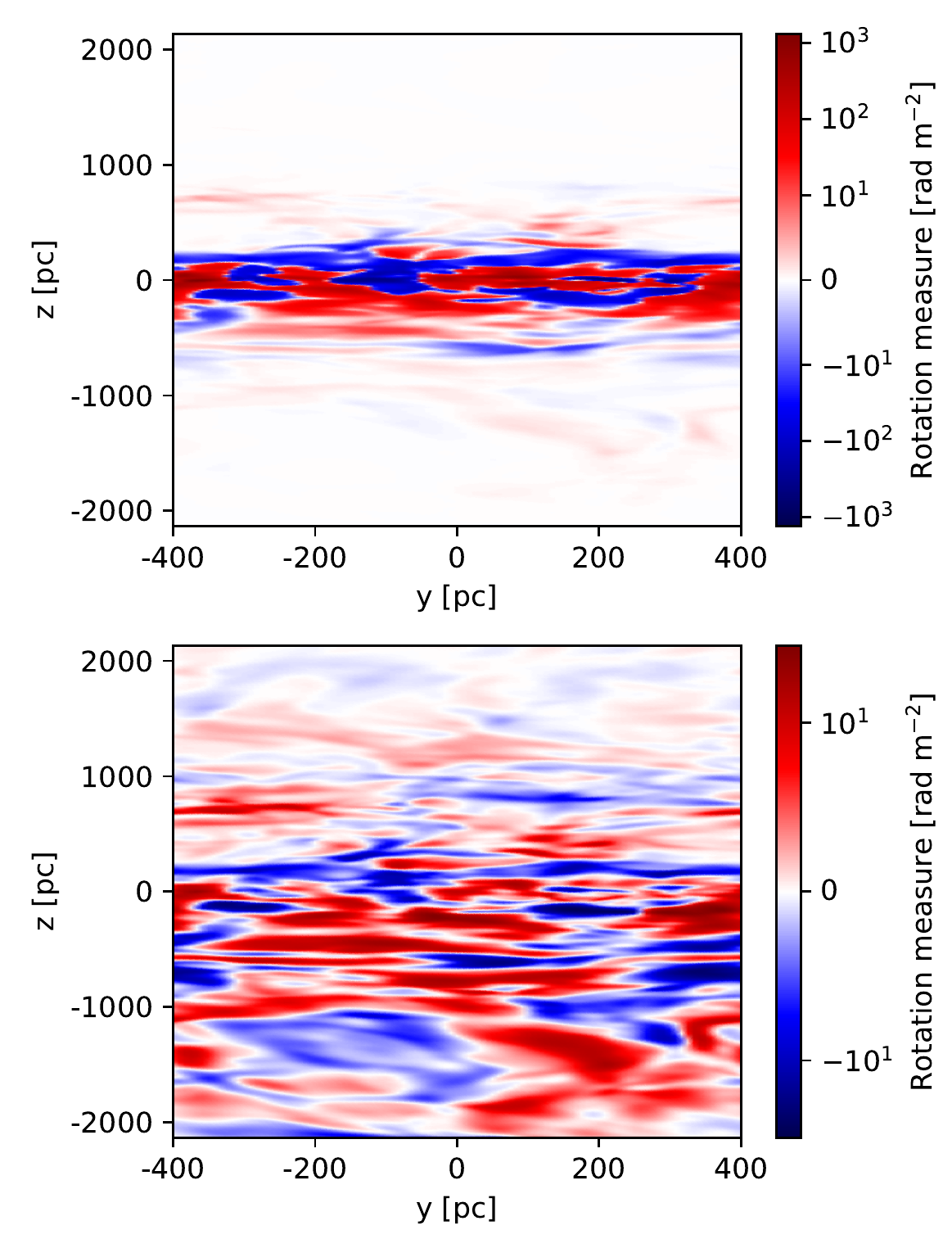}
	
    \caption{Rotation measure along the $x$ axis for Run25 at 
    $T\simeq 1508$ Myr. This time step is in the middle of the dynamical phase. 
    \textit{Top panel: } Mod1 is used, the thermal electron density is proportional to the ISM density. \textit{Bottom panel: } Mod2 is used, the thermal electron density is constant across the box.}
    
    \label{fig:RMx_appendix}
\end{figure}

\begin{figure}
	\includegraphics[width=\columnwidth]{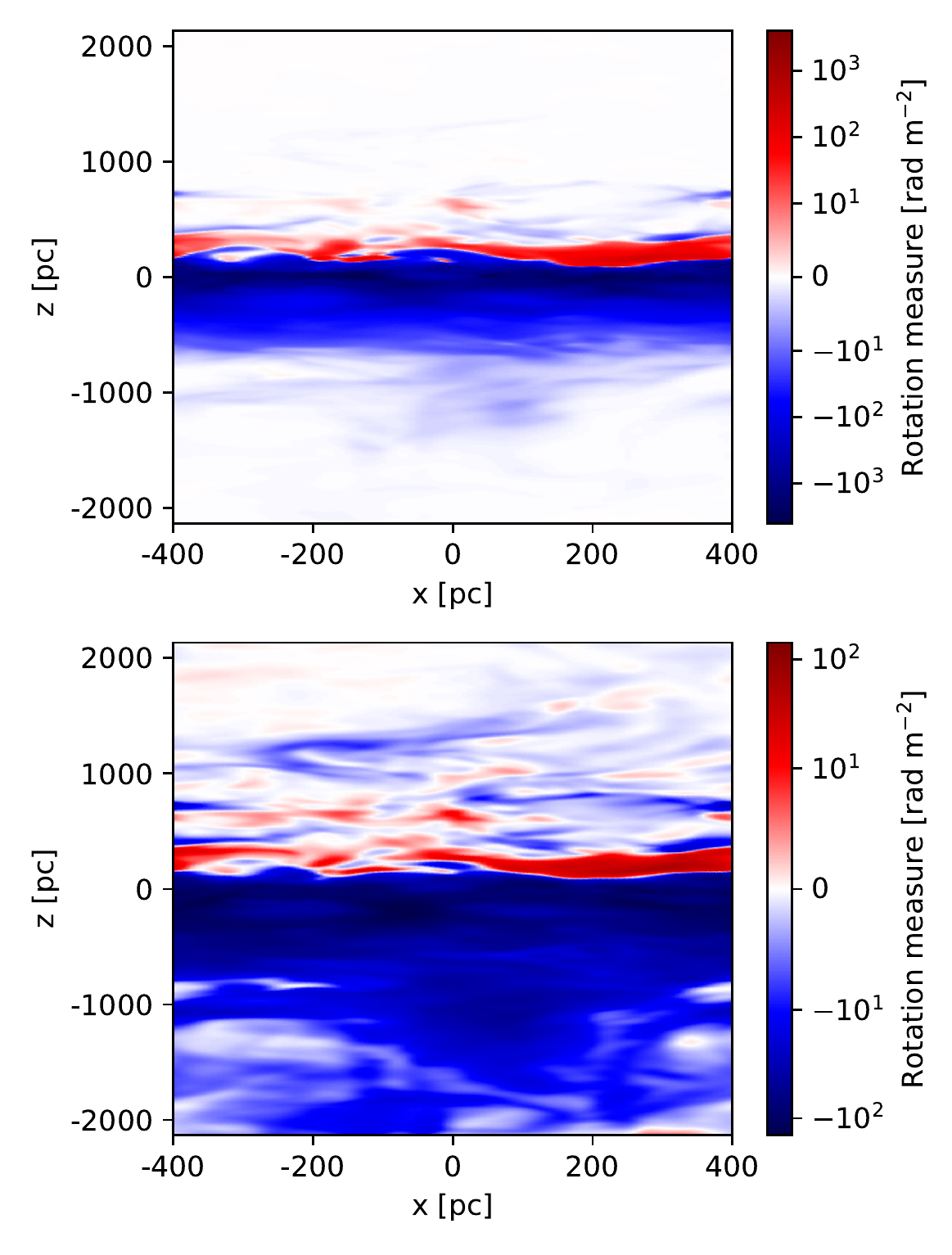}
	
    \caption{Rotation measure along the $y$ axis for Run25 at 
    $T\simeq 1508$ Myr. 
    \textit{Top panel: } Mod1 is used, the thermal electron density is proportional to the ISM density. \textit{Bottom panel: } Mod2 is used, the thermal electron density is constant across the box.}
    
    \label{fig:RMy_appendix}
\end{figure}

\begin{figure}
	\includegraphics[width=\columnwidth]{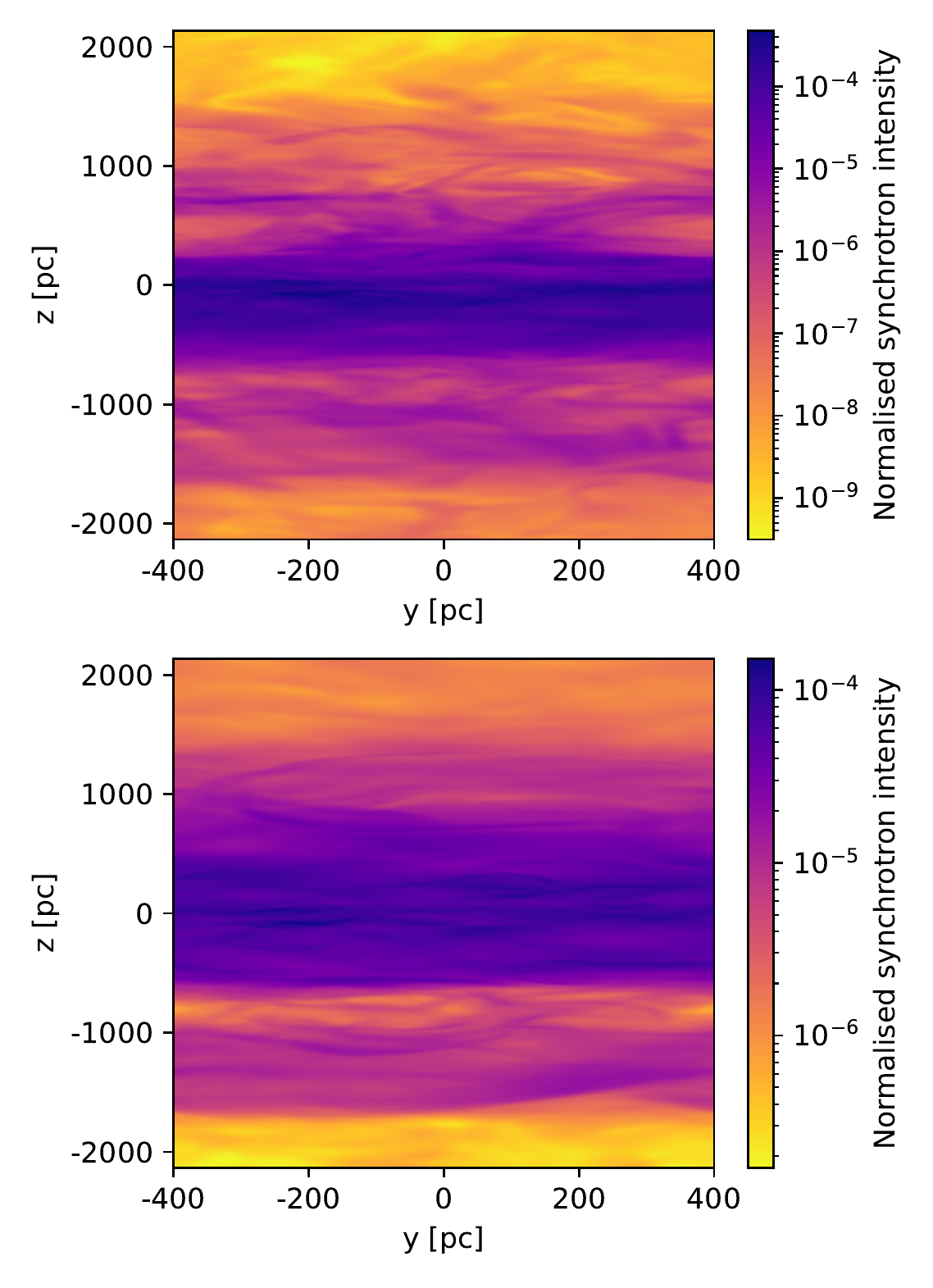}
	
    \caption{Normalised synchrotron radiation intensity along 
    the $x$ axis for Run25 at $T\simeq 1508$ Myr. This 
    time is in the middle of the dynamical phase. 
    \textit{Top panel: } CRMod1 is used, which is based 
    on equipartition between CR and magnetic energy. 
    \textit{Bottom panel: } CRMod2 is used, in which 
    the CR electrons density is constant across the 
    box.}
    
    \label{fig:Ix_appendix}
\end{figure}

\begin{figure}
	\includegraphics[width=\columnwidth]{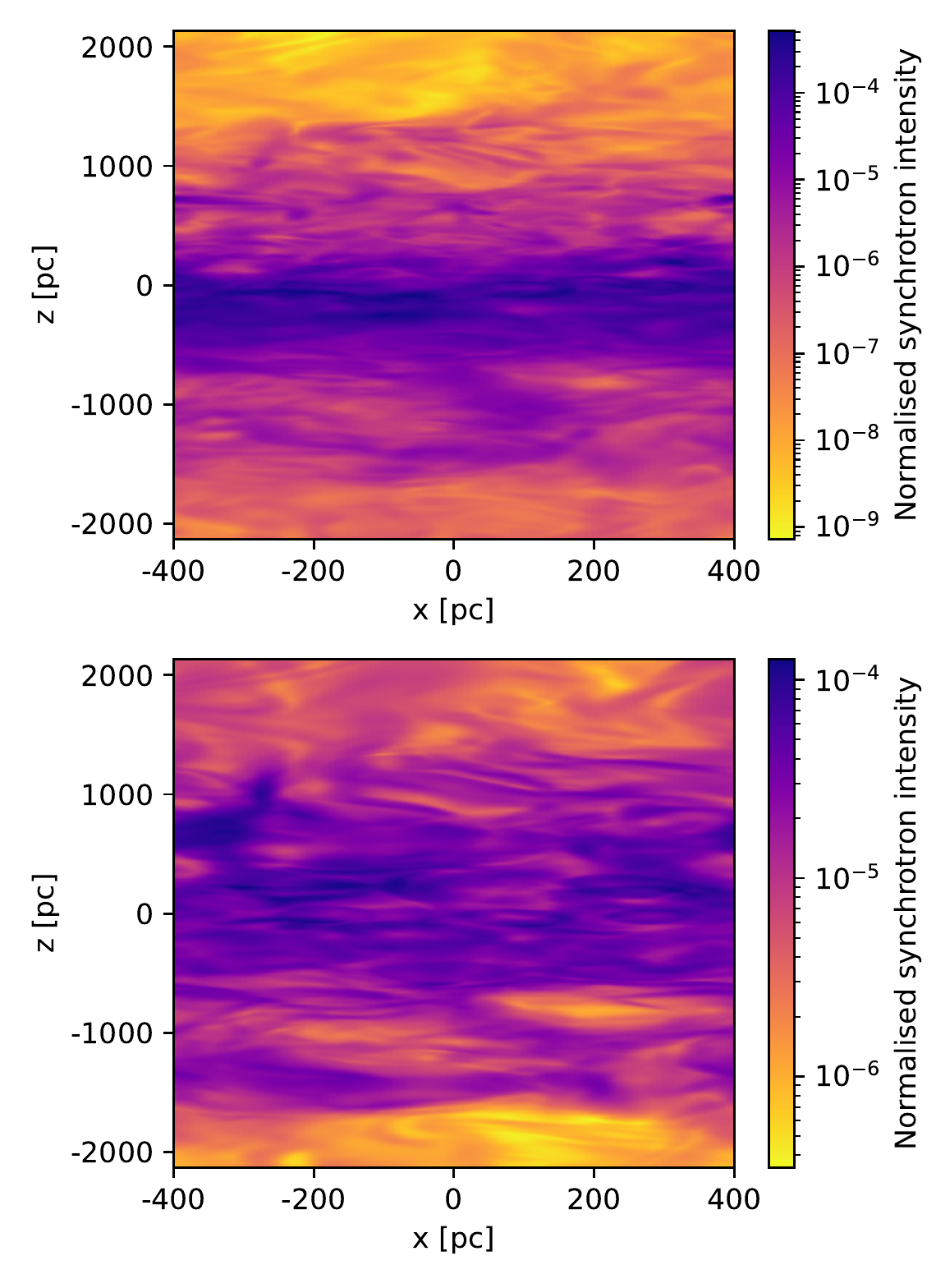}
	
    \caption{Normalised synchrotron radiation intensity along 
    the $y$ axis for Run25 at $T\simeq 1508$ Myr. This 
    time is in the middle of the dynamical phase. 
    \textit{Top panel: } CRMod1 is used, which is based 
    on equipartition between CR and magnetic energy. 
    \textit{Bottom panel: } CRMod2 is used, in which 
    the CR electrons density is constant across the 
    box.}
    
    \label{fig:Iy_appendix}
\end{figure}

\begin{figure*}
	\includegraphics[width=2\columnwidth]{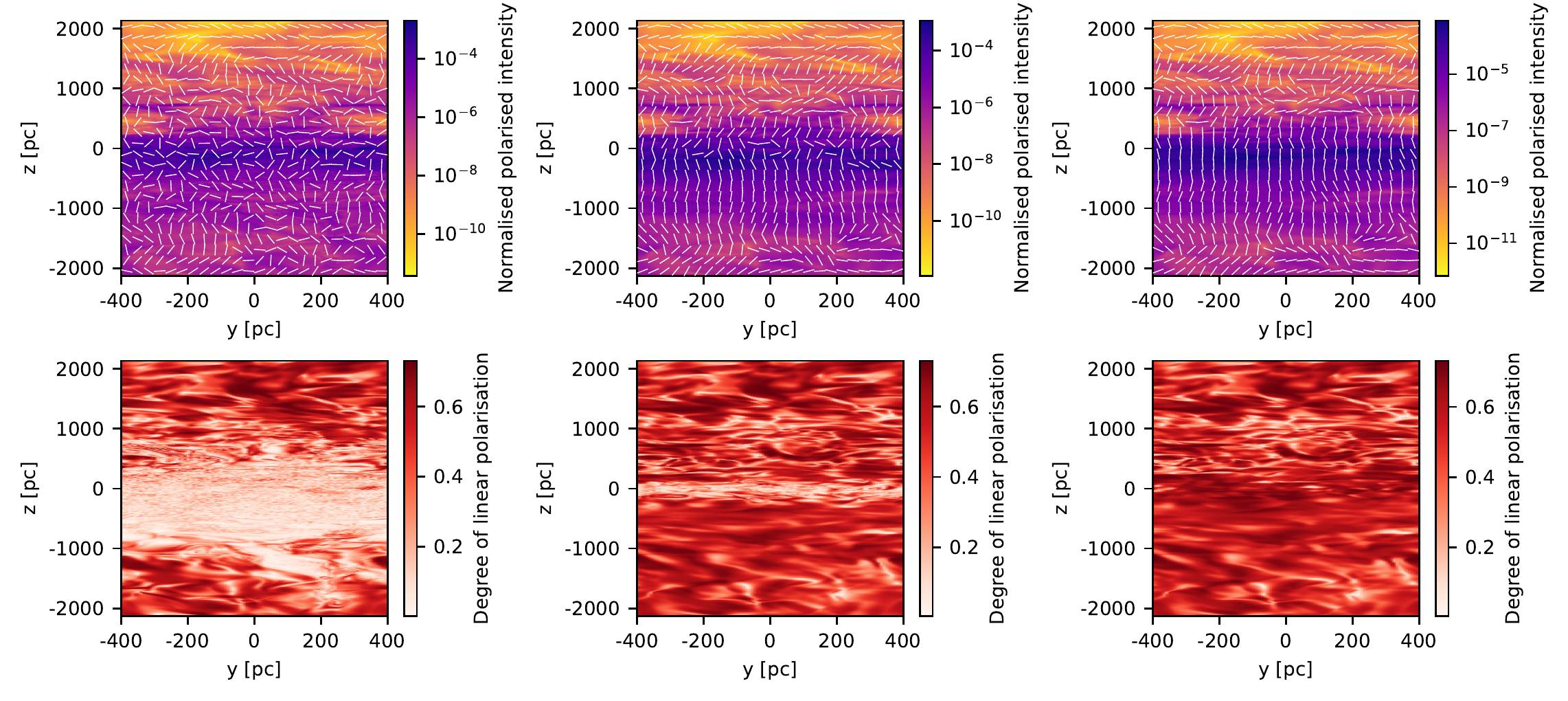}
    \caption{\textit{First row:} Normalised polarised intensity maps $\mathrm{PI}_{\nu}$ 
    with respect to the $x$ axis. The lines indicate the observed angle of polarisation $\Gamma_\nu$ with respect to the $x$ axis. \textit{Second row:} Degree of linear polarisation $p_\nu$ with respect to the $x$ axis. 
    The three columns correspond, respectively, to a frequency of observation of 
    70~MHz, 1.4~GHz, and 5~GHz for Run25 at 
    $T\simeq1508$ Myr, i.e.~in the dynamical phase. 
    We used Mod1 and CRMod1 to compute all the presented quantities.}
    \label{fig:few_maps_appendix_x}
\end{figure*}

\begin{figure*}
	\includegraphics[width=2\columnwidth]{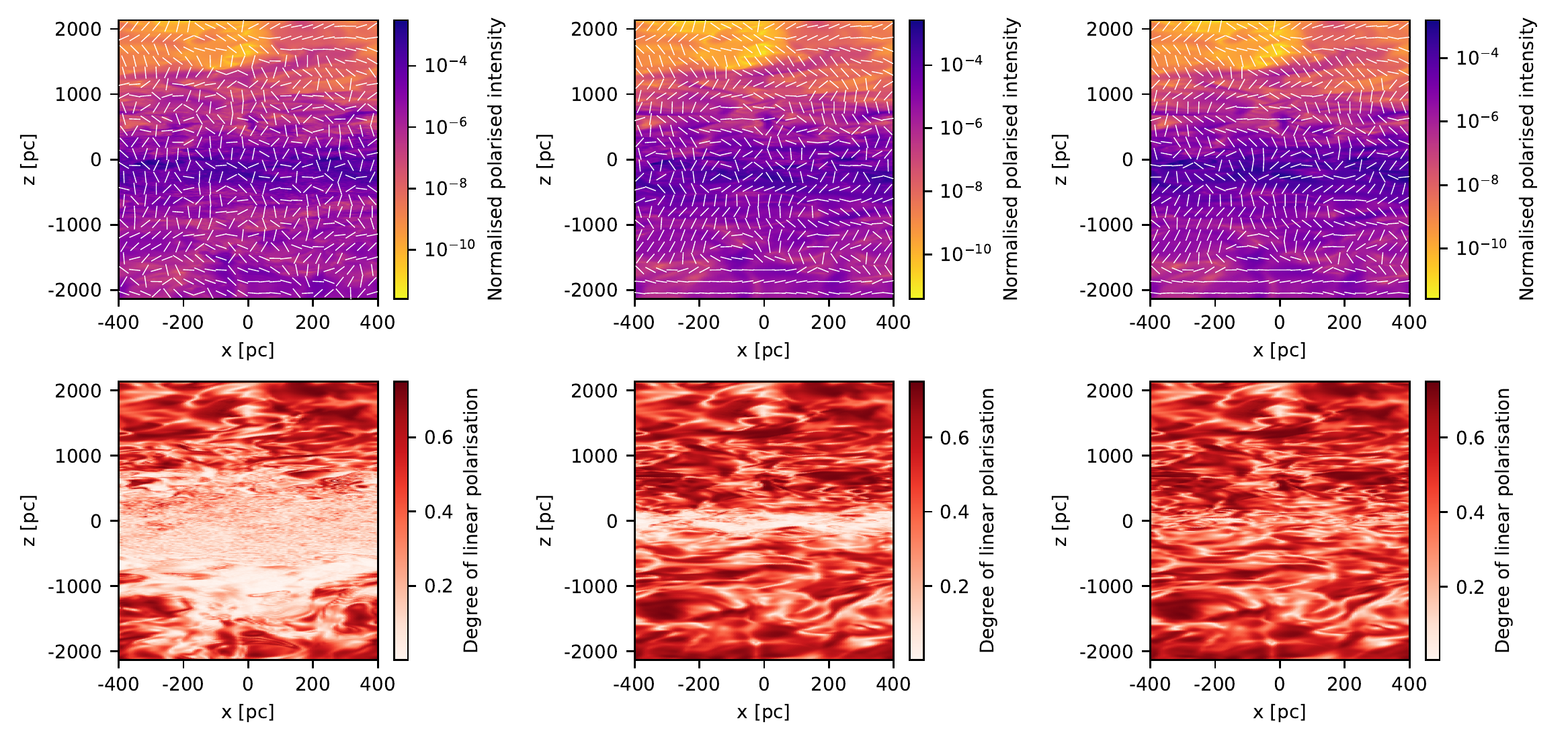}
    \caption{\textit{First row:} Normalised polarised intensity maps $\mathrm{PI}_{\nu}$ 
    with respect to the $y$ axis. The lines indicate the observed angle of polarisation $\Gamma_\nu$ with respect to the $y$ axis. \textit{Second row:} Degree of linear polarisation $p_\nu$ with respect to the $y$ axis. 
    The three columns correspond, respectively, to a frequency of observation of 
    70~MHz, 1.4~GHz, and 5~GHz for Run25 at 
    $T\simeq1508$ Myr, i.e.~in the dynamical phase. 
    We used Mod1 and CRMod1 to compute all the presented quantities.}
    \label{fig:few_maps_appendix_y}
\end{figure*}

We display in this section the results obtained for other two axes of the simulation, namely the radial $x$ and azimuthal $y$ directions. Typical correlation lengths of the RM in the dynamical can be found in Table~\ref{Table:corr_length_x_appendix} for the $x$ axis and in Table~\ref{Table:corr_length_y_appendix} for the $y$ axis. The equivalent maps to \fref{fig:RMz with 2 models} for the other two axes are shown in \fref{fig:RMx_appendix} and \fref{fig:RMy_appendix}.  The normalised synchrotron intensity along, respectively, the $x$ and $y$ directions can be found in \fref{fig:Ix_appendix} and \fref{fig:Iy_appendix}. Finally in \fref{fig:few_maps_appendix_x} and \fref{fig:few_maps_appendix_y} the six contour plots of \fref{fig:maps_1} can be found for a LOS along the radial and azimuthal directions.

\begin{table}
\centering    
\begin{tabular}{c|cccc}
\hline
& & Run25 & Run50 & Run100\\ \hline 
\\[-1em]
\multirow{2}{*}{$\ell_{\mathrm{min}}$ [pc]} & Mod1 & $54\pm8$ & $45\pm4$ & $60\pm 3$ \\
& Mod2 & $44\pm7$ & $40\pm5$ & $68\pm4$\\ \hline
\multirow{2}{*}{$\ell_{\mathrm{max}}$ [pc]} & Mod1 & $116\pm11$ & $96\pm5$ & $92\pm 6$ \\
& Mod2 & $124\pm10$ & $104\pm4$ & $103\pm5$\\ 

\end{tabular}
\caption{Comparison for $\mathrm{RM}_x$ of the
time average over the last 500 Myr of each run 
of the minimum correlation length ($\ell_\mathrm{min}$) and the maximum one ($\ell_\mathrm{max}$)
for the two models and the three simulation data sets.
The errors are taken as one standard deviation of the distribution.} 
\label{Table:corr_length_x_appendix}   
\end{table}

\begin{table}
\centering    
\begin{tabular}{c|cccc}
\hline
& & Run25 & Run50 & Run100\\ \hline 
\\[-1em]
\multirow{2}{*}{$\ell_{\mathrm{min}}$ [pc]} & Mod1 & $224\pm16$ & $235\pm7$ & $158\pm 10$ \\
& Mod2 & $355\pm2$ & $345\pm4$ & $262\pm7$\\ \hline
\multirow{2}{*}{$\ell_{\mathrm{max}}$ [pc]} & Mod1 & $355\pm11$ & $330\pm4$ & $228\pm 11$ \\
& Mod2 & $378\pm5$ & $364\pm2$ & $307\pm5$\\ 

\end{tabular}
\caption{Comparison for $\mathrm{RM}_y$ of the
time average over the last 500 Myr of each run 
of the minimum correlation length ($\ell_\mathrm{min}$) and the maximum one ($\ell_\mathrm{max}$)
for the two models and the three simulation data sets.
The errors are taken as one standard deviation of the distribution.} 
\label{Table:corr_length_y_appendix}   
\end{table}

%%%%%%%%%%%%%%%%%%%%%%%%%%%%%%%%%%%%%%%%%%%%%%%%%%

% Don't change these lines
\bsp	% typesetting comment
\label{lastpage}
\end{document}